\documentclass{article}

\usepackage{arxiv}

\usepackage[utf8]{inputenc} % allow utf-8 input
\usepackage[T1]{fontenc}    % use 8-bit T1 fonts
\usepackage{hyperref}       % hyperlinks
\usepackage{url}            % simple URL typesetting
\usepackage{booktabs}       % professional-quality tables
\usepackage{amsfonts}       % blackboard math symbols
\usepackage{nicefrac}       % compact symbols for 1/2, etc.
\usepackage{microtype}      % microtypography
\usepackage{lipsum}		% Can be removed after putting your text content
\usepackage{graphicx}
\usepackage{doi}
\usepackage{enumitem}
\hypersetup{
pdftitle={Learning ultra-compressible hyperelasticity with splines: Constitutive asymmetries and non-unique representations},
pdfsubject={cond-mat.mtrl-sci},
pdfauthor={Miguel Angel Moreno-Mateos, Simon Wiesheier, Paul Steinmann, Ellen Kuhl},
pdfkeywords={Data-adaptive methods, Constitutive modeling, Supervised learning, Automated model discovery, Extreme compressibility, Linear optimization},
}

\hypersetup{
    colorlinks=true,
    linkcolor=blue,     % Color for internal links
    filecolor=magenta,  % Color for file links
    urlcolor=black,      % Color for URLs
    citecolor=blue,    % Color for citations
}

\usepackage{booktabs}
\usepackage{amssymb}
\usepackage{mathrsfs}
\usepackage{color}
\usepackage{graphicx}
\usepackage{latexsym}
\usepackage{tabls}
\usepackage{graphicx}
\usepackage{epsfig}
\usepackage{subfigure}
\usepackage{rotating}
\usepackage{latexsym}
\usepackage[mathcal]{eucal}
\usepackage{amssymb}
\usepackage{colortbl}
\usepackage{times}
\usepackage{longtable}
\usepackage{algorithm}
\usepackage{algpseudocode}
\usepackage{psfrag}
\usepackage{tabularx}
\usepackage{epstopdf}
\usepackage{gensymb}
\usepackage{bbm}
\usepackage{setspace}
\usepackage{fullpage}
\usepackage{url}
\usepackage{calc}
\usepackage{siunitx}

\usepackage{lineno,hyperref}
\usepackage{subfigure}
\modulolinenumbers[50]
\usepackage{epstopdf}
\usepackage{array}
\usepackage{textcomp}

\usepackage{booktabs}
\usepackage{mathrsfs}
\usepackage{multirow}
\usepackage{lscape}
\usepackage{amsmath}

\usepackage{mathtools}
\usepackage{amssymb}
\usepackage{color,soul}
\usepackage{bm}

\usepackage{titlesec}
\titleformat{\paragraph}
{\normalfont\normalsize\bfseries}{\theparagraph}{1em}{}
\titlespacing*{\paragraph}
{0pt}{3.25ex plus 1ex minus .2ex}{1.5ex plus .2ex}

\usepackage{scalerel,stackengine}
\newcommand\equalhat{\mathrel{\stackon[1.5pt]{=}{\stretchto{%
    \scalerel*[\widthof{=}]{\wedge}{\rule{1ex}{3ex}}}{0.5ex}}}}
    
\usepackage{tikz}

\usepackage{makecell}
\usepackage{bbm}
\usepackage{float}

\usepackage{amsthm}
\theoremstyle{definition}
\newtheorem*{remark}{Remark}

\usepackage{lineno}

\usepackage{cleveref}
\crefname{figure}{Figure}{Figures}
\crefname{table}{Table}{Tables}
\crefname{section}{}{}
\crefname{subsection}{Section}{Sections}
\crefname{subsubsection}{Section}{Sections}
\crefname{equation}{Equation}{Equations}
\crefname{algorithm}{Algorithm}{Algorithms}
\usepackage{etoolbox}
\crefname{section}{}{}

\usepackage[font=scriptsize]{caption}
\usepackage{cleveref}
\crefname{figure}{Figure}{Figures}
\crefname{table}{Table}{Tables}
\crefname{section}{}{}
\crefname{subsection}{Section}{Sections}
\crefname{subsubsection}{Section}{Sections}
\crefname{equation}{Equation}{Equations}
\crefname{algorithm}{Algorithm}{Algorithms}
\usepackage{etoolbox}
\crefname{section}{}{}

\usepackage{tikz}

\usepackage{makecell}
\usepackage{bbm}
\usepackage{float}

\usepackage{enumitem}

\usepackage{lineno}

\modulolinenumbers[5]

\usepackage[font=scriptsize]{caption}
\usepackage{amsthm}
\theoremstyle{definition}

\usepackage{stmaryrd}

\usepackage{xurl}

\begin{document}

\title{Learning ultra-compressible hyperelasticity with splines: Constitutive asymmetries and non-unique representations}

\author{ \href{https://orcid.org/0000-0002-3476-2180}{\includegraphics[scale=0.06]{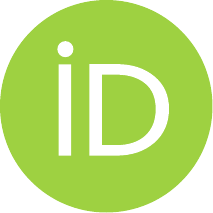}\hspace{1mm}Miguel Angel Moreno-Mateos}\thanks{Corresponding author.} \\
	Institute of Applied Mechanics\\
	Friedrich-Alexander-Universität Erlangen–Nürnberg\\
	91058, Erlangen, Germany\\
	\texttt{miguel.moreno@fau.de} \\
	%% examples of more authors
	\And
	\href{https://orcid.org/0009-0008-9625-3446}{\includegraphics[scale=0.06]{orcid.pdf}\hspace{1mm}Simon Wiesheier} \\
	Institute of Applied Mechanics\\
	Friedrich-Alexander-Universität Erlangen–Nürnberg\\
	91058, Erlangen, Germany\\
	\And
	\href{https://orcid.org/0000-0003-1490-947X}{\includegraphics[scale=0.06]{orcid.pdf}\hspace{1mm}Paul Steinmann} \\
	Institute of Applied Mechanics\\
	Friedrich-Alexander-Universität Erlangen–Nürnberg\\
	91058, Erlangen, Germany\\
	Glasgow Computational Engineering Centre\\
	University of Glasgow\\
	G12 8QQ, United Kingdom\\
	\And
	\href{https://orcid.org/0000-0002-6283-935X}{\includegraphics[scale=0.06]{orcid.pdf}\hspace{1mm}Ellen Kuhl} \\
	Department of Mechanical Engineering\\
	Stanford University\\
	318 Campus Drive, United States\\
	Institute of Applied Mechanics\\
	Friedrich-Alexander-Universität Erlangen–Nürnberg\\
	91058, Erlangen, Germany\\
}

\maketitle

\begin{abstract}
Highly compressible solids, such as foams, exhibit complex responses, including pronounced tension–compression asymmetry. Capturing such behaviors within unified hyperelastic frameworks remains challenging. 
Invariant-based hyperelastic models are commonly identified from standard tests such as homogeneous uniaxial tension/compression and simple shear, implicitly assuming a unique energy representation. Here we show that this assumption is fundamentally violated and that, oftentimes, the choice of which term should prevail is just a matter of taste.
Using spline-based strain-energy density functions as a data-adaptive tool and stress-strain experimental data for elastomeric foams, we expose this non-uniqueness, often hidden in low-parameter formulations. 
Our framework captures the volumetric deformation of ultra-light foams used in racing shoes using homogeneous experimental data from tension, compression, and shear. 
We formulate an overly rich ansatz of separable and non-separable energies in the ($\bar{I}_1$, $\bar{I}_2$, $J$) space \textit{à la} Money-Rivlin. These constructs, defined by multiplicative decompositions, resemble classical invariant-based models while generalizing them to a data-driven spline representation. This serves two purposes: (i) to capture the response under complex volumetric deformation modes and (ii) to allow non-uniqueness in the identification problem to emerge naturally.
We find that a coupling term between isochoric and volumetric deformation, such as $\Psi(\bar{I}_1,J)$ or $\Psi(\bar{I}_2,J)$, is essential and that additional coupling terms help but are not fully necessary; rather, they pronounce the non-uniqueness.
As a consequence, individual energy terms might lose physical interpretability and different models may be indistinguishable on available data. 
Importantly, these challenges are not specific to splines but extend to traditional and neural network-based models.
\end{abstract}

\keywords{Data-adaptive methods  $|$ Constitutive modeling  $|$ Supervised learning  $|$ Automated model discovery  $|$ Extreme compressibility  $|$ Linear optimization}

\section{Introduction}

How do non-separable hyperelastic energies capture asymmetries in antagonist deformation modes? Can different hyperelastic strain-energy density functions capture one observed material behavior? Should non-uniqueness degeneracies of expressive enough models be pruned and hidden or diagnosed and controlled?
Our investigation leverages a novel implementation of data-adaptive spline-based strain-energy density functions to elaborate on these questions ---in the spirit of a highly data-adaptive tool for material model calibration.

Foams are archetypal examples of highly compressible cellular soft solids \cite{Gibson1997CellularProperties}. They show a pronounced tension–compression asymmetry with a response often governed by large volumetric changes \cite{Moerman2016ControlModelling}. Classical hyperelastic models have captured this behavior with closed-form strain-energy density functions, including hyperelastic formulations \cite{Schrodt2005HyperelasticDeformations,Silber2010LargeHyperelasticity}. In the context of hyperelasticity, the constitutive response is fully characterized by a scalar strain energy density function, which can be formulated in terms of invariants of a deformation measure \cite{Holzapfel2002NonlinearScience,Agarwal2005AdmissibilityInvariants,Dammass2025}. The invariants are not only algebraic objects that emerge from the analysis of the strain tensors; they also have a natural geometric interpretation \cite{Kearsley1989,Miehe2004,Aydogdu2021}. Importantly, the use of invariants in constitutive models offers computational simplicity, in contrast to, for example, models based on principal stretch formulations. Compressible foams can be modeled in this framework, but a purely separable energy in isochoric and volumetric contributions can be too stiff to represent antagonistic deformation modes consistently; coupled volumetric effects are often necessary \cite{Fazekas2018DeterminationAccuracy,Li2022Large-deformationMaterial}. \textit{In this setting, mixed-invariant terms such as $\Psi(\bar{I}_1,J)$ and $\Psi(\bar{I}_2,J)$ are not merely mathematical embellishments: they provide a constitutive mechanism for coupling volumetric and isochoric deformation modes and thereby allow the model to express asymmetry between antagonist load paths} \cite{McCulloch2026DiscoveringShoes}.

At the same time, a central question in constitutive modeling is whether a single observed response uniquely determines a strain-energy density function. Recent data-driven hyperelasticity studies make clear that the answer is often no. Different strain-energy density functions, sometimes with very different analytical forms, can reproduce the same measured stress–strain data almost indistinguishably. This non-uniqueness has been emphasized in unsupervised discovery frameworks, where multiple admissible models fit the same dataset but differ in curvature, extrapolation, and interpretability \cite{Tac2024BenchmarkingHyperelasticity}. Oftentimes the problem is that the experimental data are not sufficiently rich to thoroughly probe localization and multiaxiality, e.g., in the invariant space for invariant-based models \cite{Flaschel2026UnsupervisedMeasurements,Avril2008OverviewMeasurements}. Biaxial characterizations of heterogeneous samples maximize the heterogeneity in the deformation fields \cite{Treloar1944Stress-strainDeformation,Esmaeili2023BiaxialReview,Moreno-Mateos2025b}. Indentation with a macroscopic tool in contact with a soft surface promises rich heterogeneous deformation modes \cite{Ashkenazi2025Indentation-basedPhantom,Shojaeifard2025HyperelasticIndentation,MorenoMateos2025-cutting}. Some methods apply LASSO sparsity regularization to remove ambiguous, unnecessary terms \cite{Flaschel2021UnsupervisedLaws}. Rather than treating this non-uniqueness as a defect to be hidden, it is more informative to expose it explicitly because it reveals which features are truly identified by the available experiments and which are not. The crux of the matter is that, \textit{oftentimes, the choice of which term should prevail is just a matter of taste}.

The present investigation leverages a new data-adaptive spline-based framework for extremely compressible solids as a high-resolution probe of this identifiability problem. In contrast to fixed low-parameter forms, spline-interpolated strain-energy density functions allow for excellent data-adaptivity. Closest precedents include the B-spline-based hyperelasticity framework of \cite{Dal2023Data-drivenMaterials}, the application to soft biological tissues \cite{Tikenogullari2023}, and the discrete data-adaptive hyperelastic approximations of the authors \cite{Wiesheier2023DiscreteFunctions}. Likewise relevant is the extension of the discrete data-adaptive idea to practical soft-material unsupervised calibration  \cite{Wiesheier2024VersatileMaterials}, the application to non-linear fracture mechanics \cite{Moreno-Mateos2025b}, and the framework DAVIS (data-adaptive spline-based dual-potential), a spline-based generalization for viscoelastic soft solids of the generalized standard materials framework \cite{Wiesheier2026Data-adaptiveSolids}. The present work extends that line of research to ultra-compressible foams and, for the first time, introduces explicit coupling terms such as $\Psi(\bar{I}_1,J)$ and $\Psi(\bar{I}_2,J)$ to analyze non-separable effects. 

This perspective also clarifies the role of model multiplicity. As shown by neural-network-based model discovery, physically admissible models can be expressive enough to fit the same material behavior in several different ways \cite{Chen2022PolyconvexApproach,Kalina2022AutomatedNetworks,Linden2023NeuralPhysics,Linka2023ADiscovery,Tac2024BenchmarkingHyperelasticity}; interestingly the work in \cite{Thakolkaran2025CanCANs} utilizes spline-like univariate functions inside a neural architecture. These approaches demonstrate that non-uniqueness degeneracy is not restricted to splines; it is intrinsic to rich constitutive model classes. The key issue is therefore not to eliminate non-uniqueness entirely, but to control it through physically meaningful constraints, report it transparently, and use it diagnostically to understand which deformation modes are needed for reliable calibration.

In this work we formulate energy representations from the most minimalist ansatz to the most expressive and ambiguous ansatz. We begin with only uncoupled terms and successively introduce additional coupling terms. We train the strain-energy density functions with tension, compression, and shear data for ultra-compressible foams from racing shoes from \cite{McCulloch2026DiscoveringShoes}. Our framework reveals when coupling terms are \textit{essential}, when they merely enrich already admissible families, and when multiple distinct energies remain \textit{indistinguishable} on the available data. This makes non-uniqueness a feature to be analyzed rather than a nuisance to be suppressed, and it positions data-adaptive spline energies as an ultra-fast, interpretable alternative to traditional phenomenological models and modern material discovery methods, such as those based on neural networks.

\section{Spline-based hyperelasticity for compressible solids}

\subsection{Continuum mechanics: Kinematics \& constitutive framework}

We let the kinematics of a hyperelastic continuum be defined by the right Cauchy-Green deformation tensor $\mathbf{C}=\mathbf{F}^\mathrm{T}\, \mathbf{F}$, with $\mathbf{F}\in \mathbb{R}^{3\times3}$ the deformation gradient. 

Following a multiplicative isochoric-volumetric decomposition $\mathbf{F}=\mathbf{F}_\mathrm{vol} \, \mathbf{F}_\mathrm{iso}$ \cite{Flory1961ThermodynamicMaterials}, the volumetric part is defined as $\mathbf{F}_\mathrm{vol}=J^{1/3} \,\mathbf{I}$ and the isochoric part as $\mathbf{F}_\mathrm{iso}=J^{-1/3} \, \mathbf{F}$. Consequently, $\bar{\mathbf{C}}=J^{-2/3} \, \mathbf{C}$ and $\mathbf{C}_\mathrm{vol}=J^{2/3} \, \mathbf{I}$.

The strain-energy density functions are constructed in terms of the first and second isochoric invariants of the right Cauchy--Green tensor, together with the Jacobian of the deformation gradient $J$, i.e.,
\begin{equation}\label{eq:isochoric_invariants}
\bar I_1 = J^{-2/3} \, I_1,
\qquad
\bar I_2 = J^{-4/3} \, I_2,
\qquad
J = \det(\mathbf{F}),
\end{equation}

\noindent with 
\begin{equation}
I_1 = \mathrm{tr}(\mathbf{C}),
\qquad
I_2 = \tfrac{1}{2}\big[(\mathrm{tr}\,\mathbf{C})^2 - \mathrm{tr}(\mathbf{C}^2)\big].
\end{equation}

In the sequel, the isochoric invariants inherit the same structure,
\begin{equation}
\bar{I}_1 = \mathrm{tr}(\mathbf{\bar{C}}),
\qquad
\bar{I}_2 = \tfrac{1}{2}\big[(\mathrm{tr}\,\mathbf{\bar{C}})^2 - \mathrm{tr}(\mathbf{\bar{C}}^2)\big].
\end{equation}

At this point, we consider a general strain energy density function
\begin{equation}
    \Psi = \Psi \left(\bar{I}_1(\bar{\mathbf{C}}),\bar{I}_2(\bar{\mathbf{C}}),J \right),
\end{equation}
which allows for both separable and non-separable (coupled) contributions.

The Piola--Kirchhoff stress tensor follows as
\begin{equation}\label{eq:secondPK_stress}
    \mathbf{S}
    =
    \mathbf{S}_\mathrm{iso}+\mathbf{S}_\mathrm{vol}
    =
    J^{-2/3} \mathbb{P}:\bar{\mathbf{S}}
    +
    J \frac{\partial \Psi}{\partial J} \mathbf{C}^{-1},
\end{equation}
where the first term represents the isochoric projection and the second term the volumetric contribution. 
Note that due to the general dependence of $\Psi$ on all invariants, these contributions are not independent; but they may depend on all crossed invariants.

The fourth-order projection operator reads
\begin{equation}
    \mathbb{P}=\mathbb{I}-\frac{1}{3}\mathbf{C}^{-1}\otimes\mathbf{C},
\end{equation}
leading to the equivalent expression \cite{Steinmann2012HyperelasticData}
\begin{equation}
    \mathbf{S}_\mathrm{iso}
    =
    J^{-2/3} \left[ \bar{\mathbf{S}} -\frac{1}{3} \left[\mathbf{C}:\bar{\mathbf{S}}\right] \mathbf{C}^{-1} \right].
\end{equation}

The fictitious stress tensor is obtained via the chain rule as
\begin{equation}\label{eq:Sbar}
    \bar{\mathbf{S}} =
    2 \left[
    \frac{\partial \Psi}{\partial \bar{I}_1} \frac{\partial \bar{I}_1}{\partial \bar{\mathbf{C}}}
    +
    \frac{\partial \Psi}{\partial \bar{I}_2} \frac{\partial \bar{I}_2}{\partial \bar{\mathbf{C}}}
    \right]
    =
    2\frac{\partial \Psi}{\partial \bar{I}_1} \mathbf{I}
    +
    2
    \frac{\partial \Psi}{\partial \bar{I}_2}
    \left[\bar{I}_1 \mathbf{I} - \bar{\mathbf{C}}\right].
\end{equation}

\subsection{B-Spline interpolated strain energy density}

To represent separable and coupled material responses, we approximate the strain energy density following the classical tradition of Rivlin-type constitutive expansions. The strain-energy density function is formulated as a truncated series in a chosen set of invariants, here $\{\bar{I}_1,\bar{I}_2,J\}$ retaining single-variable contributions and pairwise interaction terms while neglecting higher-order couplings. This is analogous in spirit to the polynomial invariant expansions of Rivlin and Saunders and to the truncated stretch-based representations later popularized by Ogden \cite{Rivlin1948,RivlinSaunders1951,ValanisLandel1967,Ogden1972,Basar2000NonlinearSolids,Bonet1997NonlinearAnalysis},
%\begin{equation}\label{eq:B-spline_energy}
%    \Psi^{(\bar{I}_1,\bar{I}_2,J)}=
%    \Psi^{(\bar{I}_1)}+
%    \Psi^{(\bar{I}_2)} +
%    \Psi^{(J)} +
%    \underbrace{g^{(J)} \, h^{(\bar{I}_1)}}_{\Psi^{(\bar{I}_1,J)}} +
%    \underbrace{i^{(J)} \,j^{(\bar{I}_2)}}_{\Psi^{(\bar{I}_2,J)}} +
%    \underbrace{k^{(\bar{I}_1)} \,l^{(\bar{I}_2)}}_{\Psi^{(\bar{I}_1,\bar{I}_2)}},
%\end{equation}
\begin{equation}\label{eq:B-spline_energy}
    \Psi^{(\bar{I}_1,\bar{I}_2,J)}=
    \sum_{\alpha \in \mathcal{A}} \Psi^{(x_\alpha)}+
    \sum_{\substack{\alpha,\beta \in \mathcal{A} \\ \alpha < \beta}} \Psi^{(x_\alpha,x_\beta)},
    \qquad
\mathcal{A}=\{1,2,3\},
\quad
(x_1,x_2,x_3)=(\bar I_1,\bar I_2,J).
\end{equation}

The first sum in~\cref{eq:B-spline_energy} yields $\Psi^{(\bar{I}_1)}$, $\Psi^{(\bar{I}_2)}$, and $\Psi^{(J)}$ terms that are univariate spline functions; whereas the second sum produces $\Psi^{(\bar{I}_1,J)}$, $\Psi^{(\bar{I}_2,J)}$, and $\Psi^{(\bar{I}_1,\bar{I}_2)}$ terms, which introduces coupling through multiplications of univariate spline functions. This construction is a particular representation of a multivariate spline, retaining expressiveness while controlling the number of unknowns.\footnote{The reader might think of a hypothetical trivariate coupling term $\Psi^{(\bar{I}_1,\bar{I}_2,J)}$ that consists of three univariate splines. For the kinematics at hand (ultra-compressible foams under homogeneous deformation), our results will show that the twofold coupling term $\Psi^{(\bar{I}_1,\bar{I}_2)}$ after calibration remains consistently inactive. This anticipates that a threefold coupling term is not necessary; we do not include it in the ansatz but we acknowledge that it might become relevant for other kinematics in future investigations.}

All spline functions are represented using cubic B-spline basis functions $\phi_i$ and corresponding control points $c_i$.\footnote{Cubic B-splines ensure $C^2$-continuity, which is required for consistent linearization in finite element implementations.} The additive contributions read
\begin{align}\label{eq:univariate_splines}
    \Psi^{(\bar{I}_1)} &= 
    \sum_{p=1}^{n_1} c^{(\bar{I}_1)}_p \, \phi_p(\bar{I}_1), \quad
    \Psi^{(\bar{I}_2)} = 
    \sum_{p=1}^{n_2} c^{(\bar{I}_2)}_p \, \phi_p(\bar{I}_2), \quad
    \Psi^{(J)} =
    \sum_{p=1}^{n_3} c^{(J)}_p \, \phi_p(J).
\end{align}

The coupling terms are constructed via multiplicative decompositions of univariate spline functions,
\begin{align}\label{eq:coupling_terms}
    \Psi^{(\bar{I}_1,J)} &=
    h^{(\bar{I}_1)} g^{(J)}
    =
    \left[ \sum_{p=1}^{n_4} c_p^{(h)} \phi_p(\bar{I}_1) \right]
    \left[ \sum_{q=1}^{n_5} c_q^{(g)} \phi_q(J) \right], \\
    \Psi^{(\bar{I}_2,J)} &=
    j^{(\bar{I}_2)} i^{(J)}
    =
    \left[ \sum_{p=1}^{n_6} c_p^{(j)} \phi_p(\bar{I}_2) \right]
    \left[ \sum_{q=1}^{n_7} c_q^{(i)} \phi_q(J) \right], \nonumber \\
    \Psi^{(\bar{I}_1,\bar{I}_2)} &=
    k^{(\bar{I}_1)} l^{(\bar{I}_2)}
    =
    \left[ \sum_{p=1}^{n_8} c_p^{(k)} \phi_p(\bar{I}_1) \right]
    \left[ \sum_{q=1}^{n_9} c_q^{(l)} \phi_q(\bar{I}_2) \right]. \nonumber
\end{align}

This multiplicative structure resembles classical invariant-based models while generalizing them to a data-driven spline representation.

\vspace{0.5em}
\noindent
\textbf{Interpolation-based parametrization.}
Instead of control points, we employ interpolation values as primary unknowns. The spline representation is therefore rewritten using interpolation basis functions $N_i$, also refered to as \textit{parameter sensitivity splines}, defined such that
\begin{equation}
    N_p(x) = \frac{\partial \Psi(x)}{\partial \theta_p},
    \qquad
    N_p(x_q)=\delta_{pq},
\end{equation}
where $\theta_p$ denotes the interpolation value at site $x_p$, and so that $N_j(x_i)=\delta_{ij}$ for $i=1,...,P$ and $\delta_{ij}$ the Kronecker delta function. The parameter sensitivity spline represents the sensitivity of the strain-energy density function with respect to the interpolation values. This leads to the following mathematically equivalent representation
\begin{align}\label{eq:B-spline_interp_basis}
\Psi(\bar{I}_1,\bar{I}_2,J) =& 
\sum_{p=1}^{n_1} \theta_p^{(\bar{I}_1)} N_p(\bar{I}_1) +
\sum_{p=1}^{n_2} \theta_p^{(\bar{I}_2)} N_p(\bar{I}_2) +
\sum_{p=1}^{n_3} \theta_p^{(J)} N_p(J) \\
&+
\left[\sum_{p=1}^{n_4} \theta_p^{(h)} N_p(\bar{I}_1)\right]
\left[\sum_{q=1}^{n_5} \theta_q^{(g)} N_q(J)\right] \nonumber \\
&+
\left[\sum_{p=1}^{n_6} \theta_p^{(j)} N_p(\bar{I}_2)\right]
\left[\sum_{q=1}^{n_7} \theta_q^{(i)} N_q(J)\right] \nonumber \\
&+
\left[\sum_{p=1}^{n_8} \theta_p^{(k)} N_p(\bar{I}_1)\right]
\left[\sum_{q=1}^{n_9} \theta_q^{(l)} N_q(\bar{I}_2)\right]. \nonumber
\end{align}

\begin{figure}[H]
    \centering
    \includegraphics[width=1.0\linewidth]{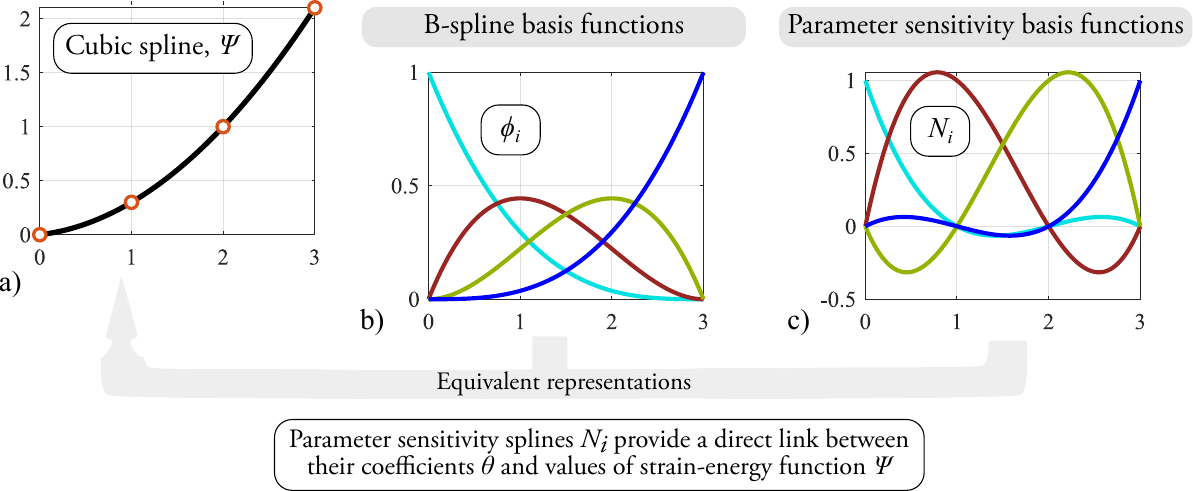}
    \caption{\textbf{Representation of strain-energy density functions with B-splines.} (a) Illustrative cubic spline utilized to model an hyperelastic strain-energy density function. Equivalent representations with (b) B-spline basis functions $\phi_i$ and control points $c_i$ as coefficients and (c) parameter sensitivity splines $N_i$ as basis functions and interpolation (collocation) points $\theta_i$ as coefficients. Note that $N_j(x_i)=\delta_{ij}$.}
    \label{fig:splines_basis}
\end{figure}

\begin{remark}
The use of interpolation values with parameter sensitivity splines (\cref{eq:B-spline_interp_basis}) and the use of control points with B-spline basis (\cref{eq:univariate_splines,eq:coupling_terms}) are two parametrizations that rely on the same underlying properties of B-spline bases. Given the interpolation sites and values, the spline construction determines the B-spline control points through a linear mapping. The relation between the control points of the splines and the interpolation values is linear and is given by the representation of the B-spline functions in \cref{eq:B-spline_energy} and \cref{eq:coupling_terms}. \cref{fig:splines_basis} provides a comparison of B-spline basis functions to parameter sensitivity basis functions; while further explanations on the relation between both parametrizations can be consulted in \cref{sec:linear_constraints}.
\end{remark}

In the present work, \textit{the use of interpolation values is particularly convenient because it provides a direct and interpretable link between the parameters and the values of the strain-energy density function at physically meaningful points in the invariant space}.

The spline-based strain-energy density functions are uniquely defined by sets of interpolation (collocation) points and their corresponding interpolation values. Specifically, interpolation points are prescribed for each invariant and for each function involved in the coupling terms, i.e., for $\bar{I}_1$, $\bar{I}_2$, and $J$, as well as for the auxiliary functions $h^{(\bar{I}_1)}$, $g^{(J)}$, $j^{(\bar{I}_2)}$, $i^{(J)}$, $k^{(\bar{I}_1)}$, and $l^{(\bar{I}_2)}$. The associated interpolation values are collected in parameter vectors:
$\boldsymbol{\theta}_{1} = [\Psi^{(\bar{I}_1)}(\bar{I}_1^{(p)})]_{p=1}^{n_1}$,
$\boldsymbol{\theta}_{2} = [\Psi^{(\bar{I}_2)}(\bar{I}_2^{(p)})]_{p=1}^{n_2}$, and
$\boldsymbol{\theta}_{\mathrm{J}} = [\Psi^{(J)}(J^{(p)})]_{p=1}^{n_3}$ for the separable terms, and analogously
$\boldsymbol{\theta}_{\mathrm{h}} = [h^{(\bar{I}_1)}(\bar{I}_1^{(p)})]_{p=1}^{n_4}$,
$\boldsymbol{\theta}_{\mathrm{g}} = [g^{(J)}(J^{(p)})]_{p=1}^{n_5}$,
$\boldsymbol{\theta}_{\mathrm{j}} = [j^{(\bar{I}_2)}(\bar{I}_2^{(p)})]_{p=1}^{n_6}$, 
$\boldsymbol{\theta}_{\mathrm{i}} = [i^{(J)}(J^{(p)})]_{p=1}^{n_7}$, 
$\boldsymbol{\theta}_{\mathrm{k}} = [k^{(\bar{I}_1)}(\bar{I}_1^{(p)})]_{p=1}^{n_8}$, and 
$\boldsymbol{\theta}_{\mathrm{l}} = [l^{(\bar{I}_2)}(\bar{I}_2^{(p)})]_{p=1}^{n_9}$ 
for the coupling contributions. These interpolation values constitute the unknown parameters of the model and fully determine the spline representation.

The stress tensor can be expressed in terms of derivatives of $\Psi$ with respect to the invariants following \cref{eq:secondPK_stress} and \cref{eq:Sbar}. For the sake of presentation, we let the stress tensor be a linear combination of the derivative of the strain-energy density functions with respect to the invariants affected by functions of the deformation gradient $\alpha(\mathbf{F})$, $\beta(\mathbf{F})$, $\gamma(\mathbf{F})$, i.e.,
\begin{equation}
    \mathbf{S} = 
    \alpha(\mathbf{F}) \, \frac{\partial \Psi}{\partial \bar{I}_1}
    + 
    \beta(\mathbf{F}) \, \frac{\partial \Psi}{\partial \bar{I}_2}
    + 
    \gamma(\mathbf{F}) \, \frac{\partial \Psi}{\partial J}.
\end{equation}

Substituting the spline representation in~\cref{eq:B-spline_interp_basis}, the stress inherits the same structure; linear dependence on parameters of separable terms and bilinear dependence on parameters of coupling terms. Explicitly, this yields
\begin{align}\label{eq:stress_splines}
&\mathbf{S}(\bar{I}_1,\bar{I}_2,J) = \nonumber \\
&\alpha(\mathbf{F}) 
\left[ 
\sum_{p=1}^{n_1} \theta_p^{(\bar{I}_1)} \frac{\partial N_p(\bar{I}_1)}{\partial \bar{I}_1}
+
\sum_{p=1}^{n_4} \sum_{q=1}^{n_5} \theta_p^{(h)} \frac{\partial N_p(\bar{I}_1)}{\partial \bar{I}_1}  \theta_q^{(g)} N_q(J)
+
\sum_{p=1}^{n_8} \sum_{q=1}^{n_9} \theta_p^{(k)} \frac{\partial N_p(\bar{I}_1)}{\partial \bar{I}_1}  \theta_q^{(l)} N_q(\bar{I}_2)
\right] 
+ \nonumber \\
&\beta(\mathbf{F})
\left[ 
\sum_{p=1}^{n_2} \theta_p^{(\bar{I}_2)} \frac{\partial N_p(\bar{I}_2)}{\partial \bar{I}_2}
+
\sum_{p=1}^{n_6} \sum_{q=1}^{n_7} \theta_p^{(j)} \frac{\partial N_p(\bar{I}_2)}{\partial \bar{I}_2}  \theta_q^{(i)} N_q(J)
+
\sum_{p=1}^{n_8} \sum_{q=1}^{n_9} \theta_p^{(k)} N_p(\bar{I}_1)  \theta_q^{(l)} \frac{\partial N_q(\bar{I}_2)}{\partial \bar{I}_2}
\right]
 + \nonumber \\
& \gamma(\mathbf{F})
\left[
\sum_{p=1}^{n_3} \theta_p^{(J)} \frac{\partial N_p(J)}{\partial J}
+
\sum_{p=1}^{n_4} \sum_{q=1}^{n_5} \theta_p^{(h)} N_p(\bar{I}_1)  \theta_q^{(g)} \frac{\partial N_q(J)}{\partial J}
+
\sum_{p=1}^{n_6} \sum_{q=1}^{n_7} \theta_p^{(j)} N_p(\bar{I}_2)  \theta_q^{(i)} \frac{\partial N_q(J)}{\partial J}
\right],
\end{align}
where each derivative follows directly from~\eqref{eq:B-spline_interp_basis}. 

This formulation highlights that: i) the separable contributions are \emph{linear} in the parameter groups $\boldsymbol{\theta}^{(\bar{I}_1)}$, $\boldsymbol{\theta}^{(\bar{I}_2)}$, and $\boldsymbol{\theta}^{(J)}$, and ii) the coupling terms are \emph{bilinear} in parameter groups $\boldsymbol{\theta}^{(h)}$, $\boldsymbol{\theta}^{(g)}$, $\boldsymbol{\theta}^{(j)}$, $\boldsymbol{\theta}^{(i)}$, $\boldsymbol{\theta}^{(k)}$, and $\boldsymbol{\theta}^{(l)}$. Importantly, the bilinear structure of the coupling terms induces a nonlinear but structured dependence of the stress on the parameter vector, which will have direct implications for the calibration procedure.

Furthermore, several requirements must be met to describe a hyperelastic material response:
\begin{itemize}
\item \textit{Objectivity}: material frame indifference of $\Psi(\mathbf{F})$ is attained by formulating the strain-energy density functions in terms of the invariants $\bar{I}_1$, $\bar{I}_2$, and $J$ of the right Cauchy-Green tensor. 

\item \textit{Stress-free reference configuration}: the first derivatives of the energy with respect to the deformation gradient vanish at the undeformed state ($\bar{I}_1=3$, $\bar{I}_2=3$, $J=1$), i.e., $\frac{\partial \Psi}{\partial \mathbf{F}}\big|_{\mathbf{F}=\mathbf{I}} = \mathbf{0}$. This is directly fulfilled due to the natural behavior of the invariants \cite{Dammass2025},
\begin{equation}
    \frac{\partial \bar{I}_1}{\partial \mathbf{F}}\bigg|_{\mathbf{F}=\mathbf I} = \mathbf{0},
    \qquad
    \frac{\partial \bar{I}_2}{\partial \mathbf{F}}\bigg|_{\mathbf{F}=\mathbf I} = \mathbf{0}.
\end{equation}

Furthermore, zero derivatives of the coupling terms with respect to $J$ at the undeformed state naturally emanates through the design feature $h^{(\bar{I}_1)}(\bar{I}_1=3)=0$ and $j^{(\bar{I}_2)}(\bar{I}_2=3)=0$, i.e.,
\begin{equation}
    %\frac{\partial \Psi^{(\bar{I}_1,J)}}{\partial J}\bigg|_{\bar{I}_1=3} = 
    h^{(\bar{I}_1)}\frac{\partial g^{(J)}}{\partial J}\bigg|_{\mathbf{F}=\mathbf{I}} = 0,
    \qquad
    %\frac{\partial \Psi^{(\bar{I}_2,J)}}{\partial J}\bigg|_{\bar{I}_2=3} = 
    j^{(\bar{I}_2)}\frac{\partial i^{(J)}}{\partial J}\bigg|_{\mathbf{F}=\mathbf{I}}=0.
\end{equation}

Therefore, we need only to enforce the following equality constraint upon construction of the spline-based strain-energy density functions
\begin{equation}
    \frac{\partial \Psi^{(J)}}{\partial J}\bigg|_{J=1} = 0.
\end{equation}

\item \textit{Energy normalization}: The condition $\Psi(\mathbf{I})=0$ is enforced by fixing the interpolation value associated with the undeformed configuration ($\bar{I}_1=3$, $\bar{I}_2=3$, and $J=1$) to zero, i.e.,
\begin{align}
\Psi^{(\bar{I}_1)}(3) = \Psi^{(\bar{I}_2)}(3)  = h^{(\bar{I}_1)}(3)=j^{(\bar{I}_2)}(3)=k^{(\bar{I}_1)}(3)=l^{(\bar{I}_2)}(3)=\Psi^{(J)}(1)=0.
\end{align}

In addition, we enforce the following normalization condition for the $J$-dependent functions in the coupling terms,
\begin{align}
g^{(J)}(1)=i^{(J)}(1)=1.
\end{align}

\item \textit{Monotonicity and convexity}: Imposed through linear inequality constraints on the spline parameters; and detailed in \cref{sec:linear_constraints}. 

\end{itemize}

\begin{remark}
    In this work, the kinematics are fully prescribed and the spline domains are known a priori. Consequently, there is no need to extrapolate the spline-based strain-energy density functions beyond their defined domain.
\end{remark}

\subsection{Parameter identification}
The identification of the interpolation values is formulated as a constrained least-squares problem. Let $\boldsymbol{\theta}$ denote the vector of spline interpolation values at prescribed knot locations. These interpolation values uniquely define the strain energy density function through spline interpolation, as defined in \cref{eq:B-spline_interp_basis}, and constitute the unknown parameters of the model.\footnote{In practice, the interpolation values are specified at the knots and mapped internally to spline control points using the \texttt{spapi} function in \textsc{Matlab}. The optimization is thus carried out in terms of interpolation values while the corresponding control points remain implicit.}

\vspace{0.5em}
\noindent
\textbf{Loss function.}
The objective is to minimize the discrepancy between model-predicted and experimentally measured stresses across multiple deformation modes. The loss function is defined as
\begin{align}\label{eq:loss_function}
\mathcal{L}(\boldsymbol{\theta}) =
& \frac{1}{n_\mathrm{ten} \, P_{11}^{\max}}
\sum_{i=1}^{n_{\mathrm{ten}}}
\left[
\left[ P_{11}(\lambda_i;\boldsymbol{\theta}) - P_{11,i} \right]^2 
+ \left[ P_{22}(\lambda_i;\boldsymbol{\theta}) \right]^2
\right]\nonumber \\
+
& \frac{1}{n_\mathrm{com} \, P_{11}^{\max}}
\sum_{i=1}^{n_{\mathrm{com}}}
\left[
\left[ P_{11}(\lambda_i;\boldsymbol{\theta}) - P_{11,i} \right]^2 
+ \left[ P_{22}(\lambda_i;\boldsymbol{\theta}) \right]^2
\right]\nonumber \\
+
& \frac{100^2}{n_\mathrm{shr} \, P_{12}^{\max}}
\sum_{i=1}^{n_{\mathrm{shr}}}
\left[
\left[ P_{12}(\gamma_i;\boldsymbol{\theta}) - P_{12,i} \right]^2
\right],
\end{align}
\noindent with $P_{11,i}$ and $P_{12,i}$ the experimental Piola stress for uniaxial and shear deformation modes, respectively; the subscript $i$ indexes points along the loading curves.

Here, uniaxial tension, compression, and shear experiments are considered. The normalization ensures a balanced contribution of each deformation mode.\footnote{The loss function assigns greater weight to the modes associated with high maximum stresses; the loss function has units of stress.} We note that the Piola stress tensor is employed because the experimental data are reported in engineering measures, i.e., referred to the undeformed configuration. Then, the stress follows from a push-forward of the second Piola--Kirchhoff stress as $\mathbf{P} = \mathbf{F}\,\mathbf{S}$.

\vspace{0.5em}
\noindent
\textbf{Optimization problem.}
The parameter vector is composed of blocks corresponding to the different spline functions,
\begin{align}\label{eq:optim_problem}
\boldsymbol{\theta} 
&= 
\big[
\boldsymbol{\theta}_1,\,
\boldsymbol{\theta}_2,\,
\boldsymbol{\theta}_\mathrm{J},\,
\boldsymbol{\theta}_\mathrm{h},\,
\boldsymbol{\theta}_\mathrm{g},\,
\boldsymbol{\theta}_\mathrm{j},\,
\boldsymbol{\theta}_\mathrm{i},\,
\boldsymbol{\theta}_\mathrm{k},\,
\boldsymbol{\theta}_\mathrm{l}
\big],
\qquad 
\boldsymbol{\theta} \in \mathbb{R}^{n_\mathrm{tot}}, \\[4pt]
\boldsymbol{\theta}^\ast 
&= 
\operatorname*{arg\,min}_{\boldsymbol{\theta}} 
\;\mathcal{L}(\boldsymbol{\theta}), \nonumber \\[6pt]
\text{subject to} \quad 
& \mathbf{A}_{\mathrm{ineq}}\, \boldsymbol{\theta} \le \mathbf{b}_{\mathrm{ineq}}, \nonumber \\
& \mathbf{A}_{\mathrm{eq}}\, \boldsymbol{\theta} = \mathbf{b}_{\mathrm{eq}}. \nonumber
\end{align}

\noindent with $n_\mathrm{tot}$ the total number of interpolation points to be determined. The inequality constraints enforce monotonicity and directional convexity of the strain-energy density function, ensuring physically meaningful material behavior (see \cref{sec:linear_constraints}).

A key feature of the proposed formulation is the structured dependence of the stress response on the parameters:
\begin{itemize}
    \item the separable energy contributions $\Psi^{(\bar{I}_1)}$, $\Psi^{(\bar{I}_2)}$, and $\Psi^{(J)}$ depend \emph{linearly} on $\boldsymbol{\theta}$,
    \item the coupling terms $\Psi^{(\bar{I}_1,J)}$, $\Psi^{(\bar{I}_2,J)}$, and $\Psi^{(\bar{I}_1,\bar{I}_2)}$ introduce a \emph{bilinear} dependence between distinct parameter blocks.
\end{itemize}

This induces a nonlinear but highly structured optimization problem. In particular, it enables the use of efficient alternating minimization strategies, where subsets of parameters are updated sequentially while keeping the others fixed. The result is a sequence of linear subproblems.

\vspace{0.5em}
\noindent
\textbf{On polyconvexity.}
Recent efforts in constitutive modeling have emphasized the importance of polyconvexity. In principle, the present framework could be extended to incorporate such constraints, for instance by enforcing positivity of the determinant of the Hessian of $\Psi$. This could be imposed weakly via additional penalty terms in the loss function, as proposed in~\cite{Klein2026NeuralApproach}. However, such constraints introduce nonlinear dependencies on $\boldsymbol{\theta}$ and would cancel the favorable linear--bilinear structure of the problem, thereby requiring fully nonlinear optimization methods. In the present work, for the non-separable energies polyconvexity is not strictly enforced; we restrict ourselves to linear constraints ensuring monotonicity and directional convexity, which already yield robust and accurate constitutive representations.

\subsection{Alternating linear least-squares optimization: Block-separable structure}

Let us recall the strain-energy density function as a construct containing additive and multiplicatively coupled spline functions, i.e.,
\begin{equation}
\Psi = \Psi_{\mathrm{add}} + h^{(\bar{I}_1)} \, g^{(J)} + i^{(\bar{I}_2)} \, j^{(J)} + k^{(\bar{I}_1)} \, l^{(\bar{I}_2)},
\qquad
\text{with}
\qquad
\Psi_{\mathrm{add}} = \Psi^{(\bar{I}_1)} + \Psi^{(\bar{I}_2)} + \Psi^{(J)}.
\end{equation}

While the full problem is bilinear in $\boldsymbol{\theta}$, it becomes strictly linear when restricted to subsets (blocks) of parameters with the remaining parameters held fixed. Specifically, when the functions $g^{(J)}$, $j^{(J)}$, and $l^{(\bar{I}_2)}$ are fixed, the stress is linear in the coefficients of $h^{(\bar{I}_1)}$, $i^{(\bar{I}_2)}$, and $k^{(\bar{I}_1)}$. Conversely, when $h^{(\bar{I}_1)}$, $i^{(\bar{I}_2)}$, and $k^{(\bar{I}_1)}$ are fixed, the stress is linear in the coefficients of $g^{(J)}$, $j^{(J)}$, $l^{(\bar{I}_2)}$. This structure is exploited using an alternating optimization scheme. At each iteration, the parameter vector is partitioned into two blocks,
\begin{align}
\text{Block 1:} & \quad \{\text{additive},\; h,\; i,\; k\}, 
\qquad &&\mathrm{with} \quad 
&&&\boldsymbol{\theta}_\mathrm{block,1} = [
\boldsymbol{\theta}_1,\,
\boldsymbol{\theta}_2,\,
\boldsymbol{\theta}_\mathrm{J},\,
\boldsymbol{\theta}_\mathrm{h},\,
\boldsymbol{\theta}_\mathrm{i},\,
\boldsymbol{\theta}_\mathrm{k}
], \\
\text{Block 2:} & \quad \{\text{additive},\; g,\; j,\; l\},
\qquad &&\mathrm{with} \quad
&&&\boldsymbol{\theta}_\mathrm{block,2} = [
\boldsymbol{\theta}_1,\,
\boldsymbol{\theta}_2,\,
\boldsymbol{\theta}_\mathrm{J},\,
\boldsymbol{\theta}_\mathrm{g},\,
\boldsymbol{\theta}_\mathrm{j},\,
\boldsymbol{\theta}_\mathrm{l}]
. \nonumber
\end{align}

Each block is optimized sequentially while keeping the complementary parameters fixed. For a given block, the residuals depend linearly on the active parameters, leading to a linear least-squares problem of the form
\begin{equation}\label{eq:min_lsq_blockproblem}
\min_{\boldsymbol{\theta}_{\text{block},i}}
\| \mathbf{A}(\boldsymbol{\theta}_{\text{block},i};\boldsymbol{\theta}_{\text{block},-i}) \, \boldsymbol{\theta}_{\text{block},i} - \mathbf{y} \|^2,
\end{equation}

\noindent subject to the linear inequality and equality constraints in \cref{eq:optim_problem}. Note that the subscript $-i$ in \cref{eq:min_lsq_blockproblem} denotes all parameter blocks other than $i$, i.e., the other block.

The matrix $\mathbf{A}$ corresponds to the sensitivity of the stress components with respect to the active parameters ($\boldsymbol{\theta}_{\mathrm{block},i}$) and fixing the parameters in the frozen block ($\boldsymbol{\theta}_{\mathrm{block},-i}$), i.e., the entries of the matrix are obtained by evaluating the derivatives of the stress with respect to the spline coefficients. Recalling the structure of the stress in \cref{eq:stress_splines}, this calculation reduces to evaluating spline basis functions (and their derivatives) at the current deformation state. In turn, the residual vector $\mathbf{y}$ contains the experimental stress data to be fitted and the zero lateral stretch condition. Both structures read
\begin{equation}
    \mathbf{A}=
    \left[
    \begin{array}{ccc}
        \frac{\partial P_{(11/12),1}}{\partial \theta_1}  & \dots & \frac{\partial P_{(11/12),1}}{\partial \theta_p} \\
        \vdots & & \vdots \\
        \frac{\partial P_{(11/12),n}}{\partial \theta_1}  & \dots & \frac{\partial P_{(11/12),n}}{\partial \theta_p} \\
        \frac{\partial P_{22,1}}{\partial \theta_1}  & \dots & \frac{\partial P_{22,1}}{\partial \theta_p} \\
        \vdots & & \vdots \\
        \frac{\partial P_{22,n}}{\partial \theta_1}  & \dots & \frac{\partial P_{22,n}}{\partial \theta_p}
    \end{array}
    \right]
    ,
    \qquad
    \mathbf{y}=
    \left[
    \begin{array}{c}
        P_{(11/12),1}^\mathrm{exp}\\
        \vdots\\
        P_{(11/12),n}^\mathrm{exp} \\
        0 \\
        \vdots \\
        0
    \end{array}
    \right]
    \quad
    \begin{array}{l}
        \makebox[3.5cm][l]{$
        \left.\vphantom{\begin{array}{c} \vdots \\ \vdots \\ \vdots \end{array}}\right\}
        \;\text{\small $P_{(11/12)}^\mathrm{exp}$ residual}$} \\
        \makebox[3.5cm][l]{$
        \left.\vphantom{\begin{array}{c} \vdots \\ \vdots \\ \vdots \end{array}}\right\}
        \;\text{\small $P_{22}=0$ residual}$}
    \end{array}
\end{equation}

\noindent for $p = \mathrm{numel}(\boldsymbol{\theta}_{\mathrm{block},i}), \; \forall \; i \in \{1,2\}$. Note that the stress $P_{11}$ is used for the rows that correspond to uniaxial deformation modes and $P_{12}$, for those corresponding to simple shear. Likewise, the rows that enforce zero lateral stretch correspond only to the uniaxial deformation modes.

The proposed alternating scheme transforms a nonlinear optimization problem into a sequence of linear least-squares problems, which can be solved efficiently and robustly. Each linear least-squares subproblem is solved using the constrained \textsc{Matlab} solver \texttt{lsqlin}. This approach significantly reduces sensitivity to initialization and improves convergence compared to fully nonlinear optimization methods.

\begin{algorithm}[H]
\caption{Alternating constrained linear least-squares optimization}
\begin{algorithmic}[1]
\Require Initial parameters $\boldsymbol{\theta}^{(0)}$, tolerance $\varepsilon$, maximum iterations $N_{\max}$
\State $n \gets 0$
\State Compute $J^{(0)} = \|\mathbf{r}(\boldsymbol{\theta}^{(0)})\|^2$

\While{$n < N_{\max}$}

    \Statex \textbf{Step 1: Update Block 1 ($\{\text{additive}, h, i, k\}$)}
    \State Assemble design matrix $\mathbf{A}_1(\boldsymbol{\theta}^{(n)}_{\text{block,2}})$ and residual vector $\mathbf{y}$
    \State Solve constrained least-squares problem
    \State $\boldsymbol{\theta}^{(n+1)}_{\text{block,1}} =
    \arg\min_{\boldsymbol{\theta}_{\text{block,1}}}
    \|\mathbf{A}_1 \boldsymbol{\theta}_{\text{block,1}} - \mathbf{y}\|^2$

    \Statex \textbf{Step 2: Update Block 2 ($\{\text{additive}, g, j, l\}$)}
    \State Assemble design matrix $\mathbf{A}_2(\boldsymbol{\theta}^{(n+1)}_{\text{block,1}})$ and residual vector $\mathbf{y}$
    \State Solve constrained least-squares problem
    \State $\boldsymbol{\theta}^{(n+1)}_{\text{block,2}} =
    \arg\min_{\boldsymbol{\theta}_{\text{block,2}}}
    \|\mathbf{A}_2 \boldsymbol{\theta}_{\text{block,2}} - \mathbf{y}\|^2$

    \State Update full parameter vector $\boldsymbol{\theta}^{(n+1)}$

    \State Compute $J^{(n+1)} = \|\mathbf{r}(\boldsymbol{\theta}^{(n+1)})\|^2$

    \If{$\dfrac{|J^{(n+1)} - J^{(n)}|}{\max(1, J^{(n)})} < \varepsilon$}
        \State \textbf{break}
    \EndIf

    \State $n \gets n + 1$

\EndWhile

\State \Return $\boldsymbol{\theta}^{(n+1)}$
\end{algorithmic}
\end{algorithm}

\section{Experimental data for ultra-compressible hyperelasticity}
We apply our spline-based framework for compressible hyperelasticity to the experimental dataset for ultra-compressible foams from running shoes reported in \cite{McCulloch2026DiscoveringShoes}. The dataset contains stress–strain data for uniaxial tension (UT), uniaxial compression (UC), and simple shear (SS). The reported stress corresponds to the engineering stress, i.e., the component $P_{11}$ of the Piola stress tensor in the uniaxial experiments and $P_{12}$ in the shear experiments. The maximum stretches are $\lambda_1 = 1.3$ for uniaxial tension, $\lambda_1 = 0.4$ for uniaxial compression, and $\gamma_{12} = 0.15$ for simple shear. Furthermore, in the simple shear experiments a constant uniaxial pre-compression of $\lambda_1 = 0.8$ is applied.

Importantly, the authors of the dataset assume zero lateral stretch, i.e., $\lambda_2 = \lambda_3 = 1$ \cite{McCulloch2026DiscoveringShoes}. This implies that the material undergoes neither lateral contraction nor expansion during tensile and compressive testing for traction-free lateral stresses ($P_{22} = P_{33} = 0$). Although we believe that this zero-Poisson’s-ratio assumption should be revisited with greater accuracy, we adopt the same kinematic assumptions, which correspond to a purely volumetric deformation ---in our view an interesting kinematic setup. The kinematics of the problem are therefore fully prescribed and are described next.

\subsection{Uniaxial tension and compression}

Consider the deformation gradient
\begin{equation}
\mathbf{F}_{\mathrm{UT}} = \mathbf{F}_{\mathrm{UC}} \equalhat \mathrm{diag}(\lambda_1, 1, 1),
\end{equation}

with stretch $\lambda > 0$. The Jacobian is $J = \lambda_1$ and the isochoric right Cauchy--Green tensor reads
\begin{equation}
\mathbf{\bar{C}} = J ^{-2/3} \mathbf{C} \equalhat \mathrm{diag}(\lambda_1^{4/3}, \lambda_1^{-2/3}, \lambda_1^{-2/3}).
\end{equation}

The corresponding invariants are
\begin{equation}
\bar I_1(\lambda_1) = \lambda_1^{-2/3} \left[\lambda_1^{2}+2\right],
\qquad
\bar I_2(\lambda_1) =  \lambda_1^{-4/3} \left[2\lambda_1^{2}+1\right],
\qquad
J(\lambda_1) = \lambda_1.
\end{equation}

Thus, uniaxial loading defines a curve $(\bar I_1(\lambda_1), \bar I_2(\lambda_1), J(\lambda_1))$ in invariant space with parameter $\lambda$.

\subsection{Simple shear with fixed axial stretch}

Consider next a simple shear deformation with fixed axial stretch $\lambda_1=0.8$,
\begin{equation}
\mathbf{F}_\mathrm{SS} \equalhat
\begin{bmatrix}
\lambda_1 & \gamma_{12} & 0 \\
0 & 1 & 0 \\
0 & 0 & 1
\end{bmatrix},
\end{equation}
where $\gamma_{12}$ is the shear parameter. The Jacobian is $J = \lambda_1$ and the isochoric right Cauchy--Green tensor is
\begin{equation}
\mathbf{\bar{C}} = J^{-2/3}\mathbf{C} \equalhat
\lambda_1^{-2/3} \begin{bmatrix}
\lambda_1^2 & \lambda_1 \gamma_{12} & 0 \\
\lambda_1 \gamma_{12} & 1 + \gamma_{12}^2 & 0 \\
0 & 0 & 1
\end{bmatrix}.
\end{equation}

The invariants are obtained from \ref{eq:isochoric_invariants} as
\begin{equation}
\bar I_1(\gamma_{12})
=
\lambda_1^{-2/3} \left[\lambda_1^2 + \gamma_{12}^2 +2 \right],
\qquad
\bar I_2(\gamma_{12})
=
\lambda_1^{-4/3} \left[2\lambda_1^2 + \gamma_{12}^2 +1 \right], 
\qquad
J = \lambda_1.
\end{equation}

Consequently, simple shear defines the curve $(\bar I_1(\lambda_1), \bar I_2(\lambda_1), J)$ with parameter $\gamma_{12}$.

\begin{figure}[h!]
    \centering
    \includegraphics[width=1.0\linewidth]{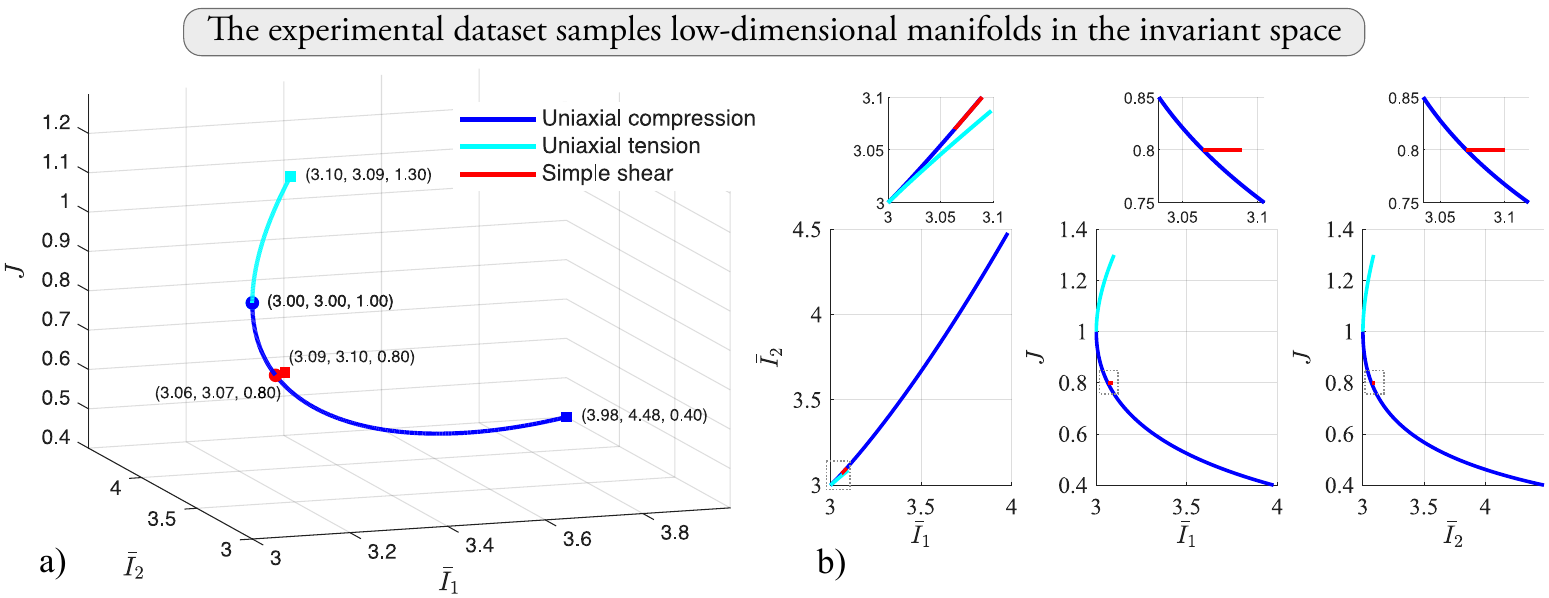}
    \caption{\textbf{Admissible invariant space for tensile and compression volumetric and shear deformation.} (a) Three-dimensional parametric curves represent the space admitted in the $(\bar I_1, \bar I_2, J)$ domain. (b) Two-dimensional projections of the curves in the $\bar I_1-\bar I_2$, $\bar I_1-J$, and $\bar I_2-J$ spaces. Insets show zoom details of the curves.}
    \label{fig:space_probed}
\end{figure}

The key question is: \textit{How sparse of a subspace are we actually sampling?} Both deformation modes are defined by one-dimensional curves embedded in the three-dimensional invariant space $(\bar I_1, \bar I_2, J)$. Simple shear evolves on a plane of constant $J$, while uniaxial loading couples all three invariants through a nonlinear relation. This observation is insightful to the identifiability discussion: standard experimental datasets do not sample a finite volume in invariant space, but rather a collection of low-dimensional manifolds. The curves are illustrated in \cref{fig:space_probed}.

\section{Results}

We explore part of the optimization landscape by employing \textbf{random initial guesses} for all spline-based terms. Specifically, the interpolation values are initialized using the \texttt{randn(1, n)} function in \textsc{Matlab}, which generates a vector of $n$ independent random numbers drawn from a standard normal distribution (zero mean and unit variance).

\subsection{Single-invariant energies $\Psi^{(\bar{I}_1)}$, $\Psi^{(\bar{I}_2)}$, and $\Psi^{(J)}$ cannot capture foam tension--compression asymmetry}\label{sec:only_uncoupled}

We begin with the simplest ansatz: purely separable energy contributions, 
$\Psi^{(\bar{I}_1,\bar{I}_2,J)} = \Psi^{(\bar{I}_1)} + \Psi^{(\bar{I}_2)} + \Psi^{(J)}$. 
For this choice, the problem is directly linear in the optimization parameters 
$\boldsymbol{\theta} = [\boldsymbol{\theta}_1, \boldsymbol{\theta}_2, \boldsymbol{\theta}_\mathrm{J}]$. 
Due to the simplicity of this ansatz, the resulting solution is unique and independent of the initial guess; a conclusion reached by performing the training for different random initial guesses.

\begin{figure}[h!]
    \centering
    \includegraphics[width=1.0\linewidth]{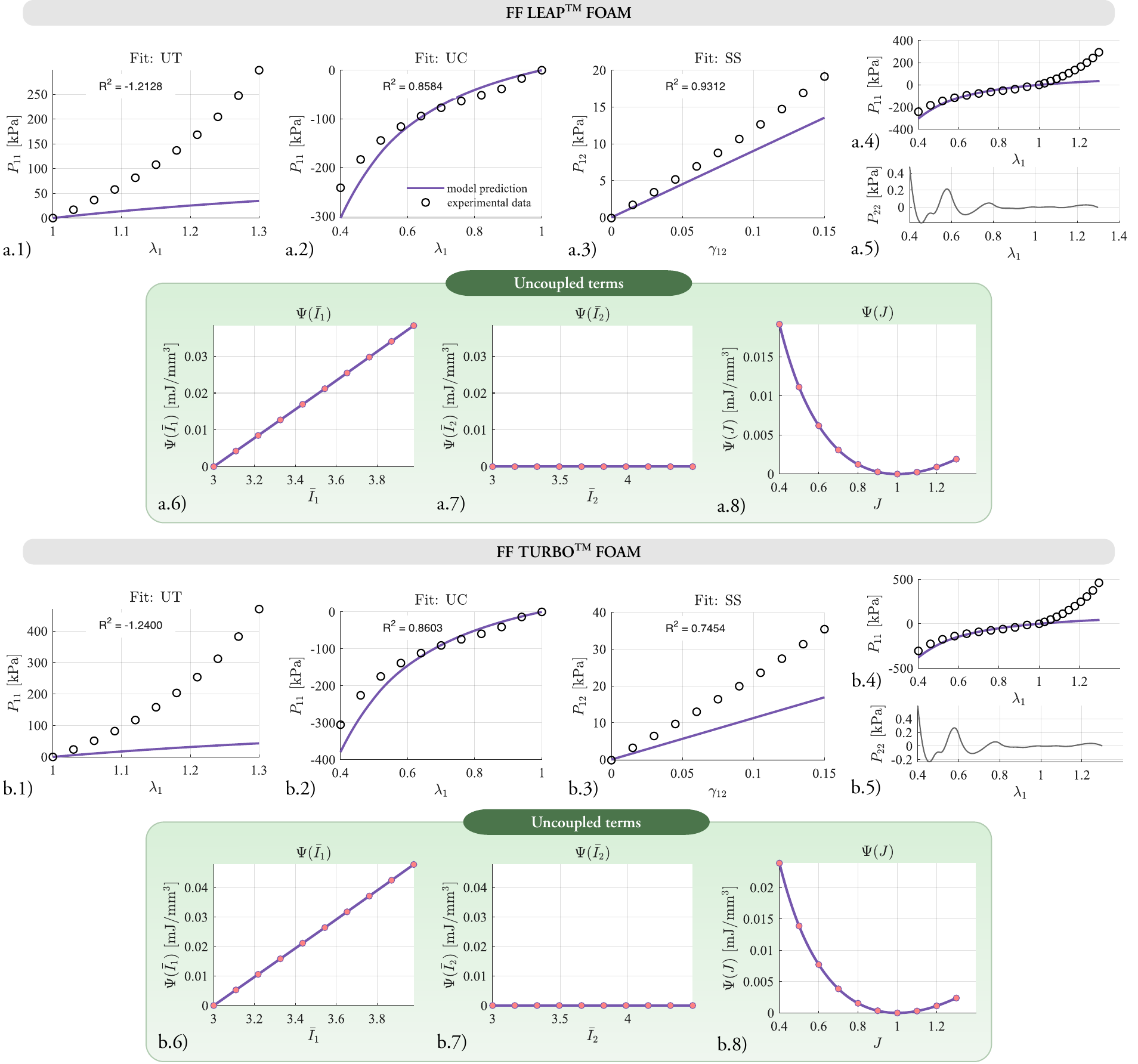}
    \caption{\textbf{Discovered hyperelastic strain-energy density functions with single-invariant terms for FF LEAP$^\mathrm{TM}$ and FF TURBO$^\mathrm{TM}$ PLUS.}
(a,b.1--4) Model predictions for uniaxial tension (UT), uniaxial compression (UC), and simple shear (SS) deformation modes of the FF LEAP$^\mathrm{TM}$ and FF TURBO$^\mathrm{TM}$ foams. Nominal stress--stretch responses are shown alongside the experimental data used in the loss function.
(a,b.5) Predicted lateral stress response, for zero lateral stress condition enforced.
(a,b.6--8) Learned spline-based strain-energy density contributions.}
    \label{fig:results_uncoupled}
\end{figure}

~\cref{fig:results_uncoupled} shows the results for both FF LEAP$^\mathrm{TM}$ and FF TURBO$^\mathrm{TM}$ foams. 
Interestingly, the identified energy is dominated by the contributions $\Psi(\bar{I}_1)$ and $\Psi(J)$, while the $\bar{I}_2$-dependent term is effectively inactive. 
Moreover, the dependence on $\bar{I}_1$ appears nearly linear over the investigated deformation range.
The predicted stresses under uniaxial tension, uniaxial compression, and simple shear are compared against the experimental data. The overall quality of the fit is unsatisfactory for both materials, with negative $\mathrm{R}^2$ values observed in some deformation modes. 
For the FF LEAP$^\mathrm{TM}$ foam $\mathrm{R}^2=\qty{-1.21}{}$ for tension, $\mathrm{R}^2=\qty{0.86}{}$ for compression, and $\mathrm{R}^2=\qty{0.93}{}$ for shear, and for the FF TURBO$^\mathrm{TM}$ foam $\mathrm{R}^2=\qty{-1.24}{}$ for tension, $\mathrm{R}^2=\qty{0.86}{}$ for compression, and $\mathrm{R}^2=\qty{0.75}{}$ for shear. 
This is expected because the assumed energy decomposition into purely separable contributions in $\bar{I}_1$, $\bar{I}_2$, and $J$ is too restrictive to capture the pronounced tension–compression asymmetry characteristic of these foams. 

The solution reflects a projection of the data onto a severely restricted functional subspace and, consequently, does not exploit the full invariant space available to the model. A natural extension is therefore to introduce a coupling term between $\bar{I}_1$ and $J$, which enables interaction between isochoric and volumetric deformation modes.

\subsection{A coupling term $\Psi^{(\bar{I}_1,J)}$ captures extreme tension--compression asymmetry}

We continue with a more elaborated ansatz: purely separable energy contributions plus a term coupling $\bar{I}_1$ and $J$, 
$\Psi^{(\bar{I}_1,\bar{I}_2,J)} = \Psi^{(\bar{I}_1)} + \Psi^{(\bar{I}_2)} + \Psi^{(J)} + \Psi^{(\bar{I}_1,J)}$. 
For this choice, the problem becomes bilinear in the optimization parameters, i.e., $\boldsymbol{\theta} = [\boldsymbol{\theta}_1, \boldsymbol{\theta}_2, \boldsymbol{\theta}_\mathrm{J}, \boldsymbol{\theta}_\mathrm{h}, \boldsymbol{\theta}_\mathrm{g}]$. 
Due to the yet limited expressive capability of this ansatz, the resulting solution is independent of the initial guess. 

\begin{figure}[h!]
    \centering
    \includegraphics[width=0.87\linewidth]{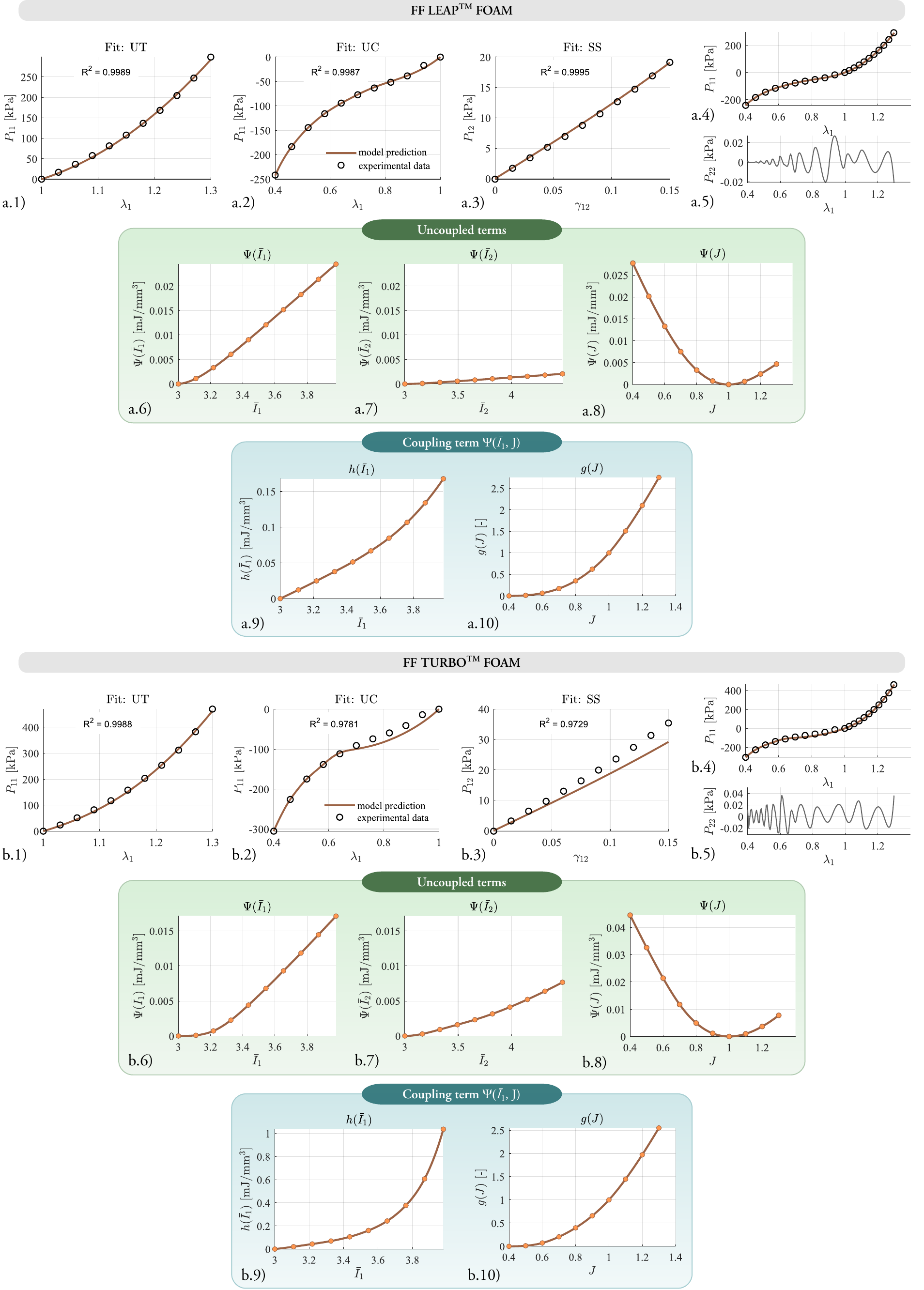}
    \caption{\textbf{Discovered hyperelastic strain-energy density functions with a mixed-invariant term $\Psi^{(\bar{I}_1,J)}$ for FF LEAP$^\mathrm{TM}$ and FF TURBO$^\mathrm{TM}$ PLUS.}
(a,b.1--4) Model predictions for uniaxial tension (UT), uniaxial compression (UC), and simple shear (SS) deformation modes of the FF LEAP$^\mathrm{TM}$ and FF TURBO$^\mathrm{TM}$ foams. Nominal stress--stretch responses are shown alongside the experimental data used in the loss function.
(a,b.5) Predicted lateral stress response, for zero lateral stress condition enforced.
(a,b.6--10) Learned spline-based strain-energy density contributions.}
    \label{fig:results_I1J}
\end{figure}

\cref{fig:results_I1J} contains the results for both foams. 
The coupling term $\Psi(\bar{I}_1,J)$ with non-separable $\bar{I}_1$-- and $J$--dependent energies plays an important role: Part of the energy originally in the uncoupled term $\Psi^{(\bar{I}_1)}$ is now accommodated in this multiplicative term (see \cref{fig:results_I1J}.9-10, a) and b), for both foams). Interestingly, both $h^{(\bar{I}_1)}$ and $g^{(J)}$ spline-based strain-energy density functions are highly non-linear, convex functions. Furthermore, the uncoupled $\Psi^{(\bar{I}_2)}$ term is in this case activated, more for the TURBO$^\mathrm{TM}$ foam than for the LEAP$^\mathrm{TM}$ one; yet its contribution is smaller than the $\Psi^{(\bar{I}_1)}$ term. A direct comparison can be made to the terms calibrated in \cite{McCulloch2026DiscoveringShoes} for the same dataset with a Neural Network: The authors restricted the training to a pre-defined potential dependence on $J$ and linear dependence on $\bar{I}_1$; i.e., $w_{11} J^{w_{11}^*}\left[\bar{I}_1-3\right]$, with $w_{11}$ and $w_{11}^*$ weights to be trained. In contrast, our method provides extreme data-adaptivity: Cubic splines are expressive to adapt in their shape and curvature. Although $g^{(J)}$ is not enforced to be monotonous, it is discovered as a monotonous function in both compression ($J<1$) and tension ($J>1$) regimes. Simultaneously, $g^{(J)}$ modulates (via multiplicative coupling) the energy function $h^{(\bar{I}_1)}$, also highly non-linear (convex). Our spline-based constructs produce a convincingly good prediction of the experimental curves. 

The overall quality of the fit is good for both foams, and especially good for the FF LEAP$^{\mathrm{TM}}$ foam. For the LEAP$^{\mathrm{TM}}$ foam (see \cref{fig:results_I1J}.a.1-3) the $\mathrm{R}^2$ value is $\mathrm{R}^2=\qty{1.00}{}$ for tension, $\mathrm{R}^2=\qty{1.00}{}$ for compression, and $\mathrm{R}^2=\qty{1.00}{}$ for shear. For the TURBO$^{\mathrm{TM}}$ foam, the fitting is slightly worse, with $\mathrm{R}^2= \qty{1.00}{}$ for tension, $\mathrm{R}^2=\qty{0.98}{}$ for compression, and $\mathrm{R}^2=\qty{0.97}{}$ for shear. Nonetheless, the reader has to recall that the fitting also includes $P_{22}=0$, and this is enforced with more prominence: a rather strong penalization in the loss function (see the 100-times larger weight in \cref{eq:loss_function}). Note that we design the loss function in this way because we believe that our spline-based energy should unconditionally represent vanishing negligible lateral stress. Thus, the lateral stress satisfactorily becomes almost zero; it is less than \qty{0.01}{\%} the maximum stresses in the deformation modes. Overall, the data-adaptive spline-based strain-energy density functions yield satisfactory predictions.

\subsection{A coupling term $\Psi^{(\bar{I}_2,J)}$ can also capture extreme volumetric deformations and reveals the ambiguity between $\bar{I}_1$ and $\bar{I}_2$}
One question remains: Can a single coupling term $\Psi^{(\bar{I}_2,J)}$ capture the tension--compression asymmetry as well? We replace the coupling term $\Psi^{(\bar{I}_1,J)}$ by a coupling term $\Psi^{(\bar{I}_2,J)}$; the ansatz now reads: $\Psi^{(\bar{I}_1,\bar{I}_2,J)} = \Psi^{(\bar{I}_1)} + \Psi^{(\bar{I}_2)} + \Psi^{(J)} + \Psi^{(\bar{I}_2,J)}$. Again, this representation converges to the same solution for different initial random guesses given the yet limited expressiveness of the ansatz.

\begin{figure}[h!]
    \centering
    \includegraphics[width=0.87\linewidth]{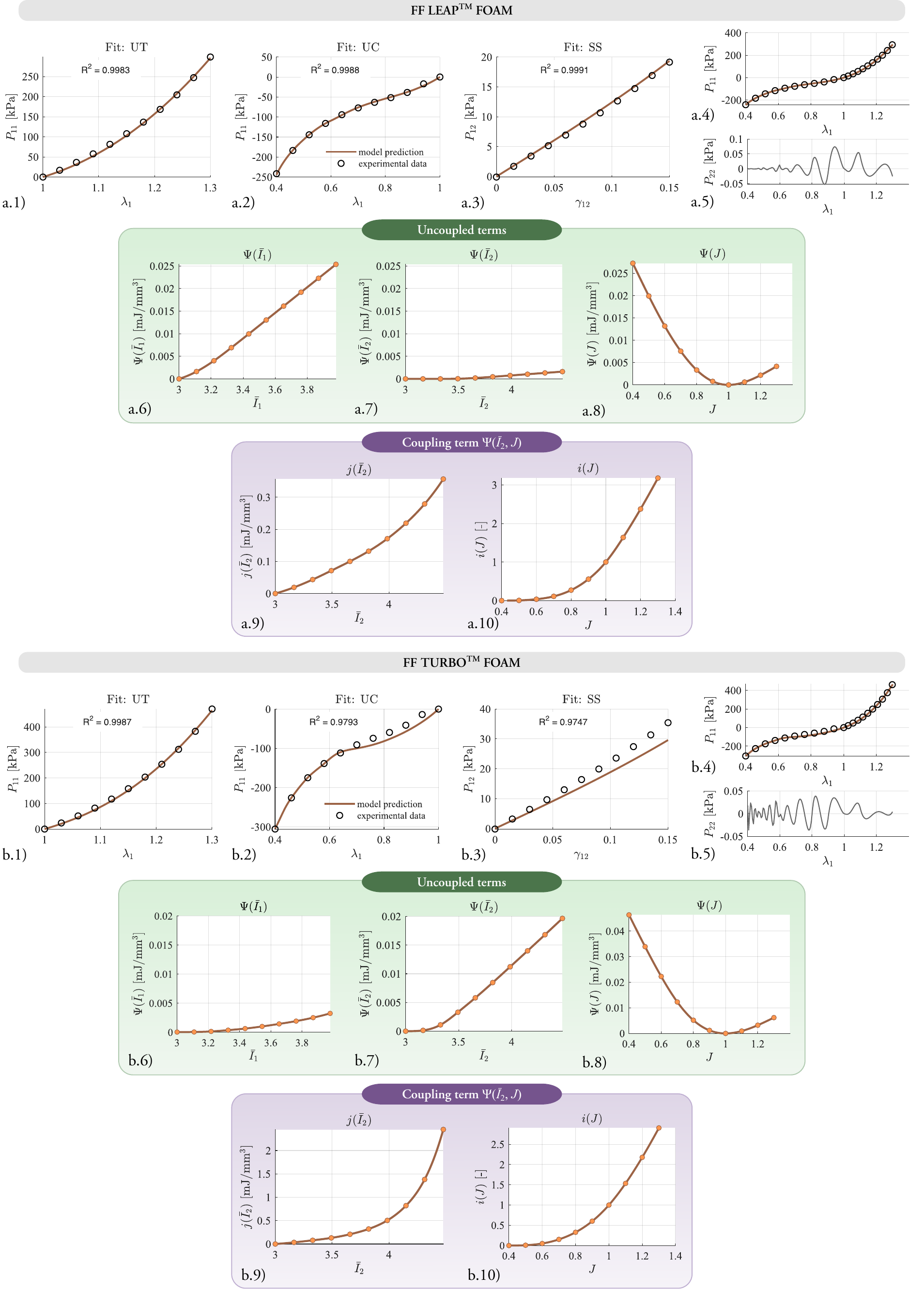}
    \caption{\textbf{Discovered hyperelastic strain-energy density functions with a mixed-invariant term $\Psi^{(\bar{I}_2,J)}$ for FF LEAP$^\mathrm{TM}$ and FF TURBO$^\mathrm{TM}$ PLUS.}
(a,b.1--4) Model predictions for uniaxial tension (UT), uniaxial compression (UC), and simple shear (SS) deformation modes of the FF LEAP$^\mathrm{TM}$ and FF TURBO$^\mathrm{TM}$ foams. Nominal stress--stretch responses are shown alongside the experimental data used in the loss function.
(a,b.5) Predicted lateral stress response, for zero lateral stress condition enforced.
(a,b.6--10) Learned spline-based strain-energy density contributions.}
    \label{fig:results_I2J}
\end{figure}

The answer is yes: A single coupling term combining $\bar{I}_2$ and $J$ provides the same good fitting of the three deformation modes and vanishing lateral stress. \cref{fig:results_I2J} contains the results: The $\mathrm{R}^2$ values are $\mathrm{R}^2=\qty{1.00}{}$ for tension, $\mathrm{R}^2=\qty{1.00}{}$ for compression, and $\mathrm{R}^2=\qty{1.00}{}$ for shear for the FF LEAP$^\mathrm{TM}$ foam; and $\mathrm{R}^2=\qty{1.00}{}$ for tension, $\mathrm{R}^2=\qty{0.98}{}$ for compression, and $\mathrm{R}^2=\qty{0.97}{}$ for shear for the FF TURBO$^\mathrm{TM}$ foam. This is convincingly similar to the previous results for the $\bar{I}_1-J$-coupling term. The similarity between $\bar{I}_1$ and $\bar{I}_2$, as revealed by the $\lambda_1$-based parametrization of the invariants in \cref{fig:space_probed}.b, further supports this finding: For small compression/tension volumetric deformations, both invariants take similar values. A consequence is that, for the kinematics at hand (ultra-compressible foams under homogeneous deformation), the $\bar{I}_1$-- and $\bar{I}_2$--based coupling terms are ambiguous and the choice of which term should prevail is just a matter of taste.

When it comes to the energy terms one interesting difference for the FF TURBO$^\mathrm{TM}$ foam is that the uncoupled energy $\Psi^{(\bar{I}_2)}$ now dominates over $\Psi^{(\bar{I}_1)}$; the latter is about \qty{10}{\%} the former (\cref{fig:results_I2J}.b.7). This is reasonable because the continuum modeling we are doing can not distinguish whether the volumetric deformation is driven by \textit{line}-element-like ($\bar{I}_1$-related) or \textit{area}-like ($\bar{I}_2$-related) deformations \cite{Kuhl2024IInvariant}. Then, a pertinent question is: Can an ansatz without the uncoupled term $\Psi^{(\bar{I}_2)}$, i.e., $\Psi^{(\bar{I}_1,\bar{I}_2,J)} = \Psi^{(\bar{I}_1)} + \Psi^{(J)} + \Psi^{(\bar{I}_2,J)}$, represent the same good predictions? We present the results for this additional case in \cref{fig:results_I2J_withoutI2}; the answer is yes. The just described behavior already shows certain ambiguity and non-uniqueness in the representation of the energy density function to yield similar good predictions.

\subsection{A coupling term $\Psi^{(\bar{I}_1,\bar{I}_2)}$ naturally remains inactive}

For completeness, we now explore the $\Psi^{(\bar{I}_1,\bar{I}_2)}$ coupling term. It is a term that only couples isochoric invariants, irrespective of the volumetric deformation. The ansatz now reads:
$\Psi^{(\bar{I}_1,\bar{I}_2,J)} = \Psi^{(\bar{I}_1)} + \Psi^{(\bar{I}_2)} + \Psi^{(J)} + \Psi^{(\bar{I}_1,\bar{I}_2)}$. One would expect this term to be relevant in deformations where there is an interaction between distortional deformation modes, i.e., the deformation cannot be decomposed into independent contributions of $\bar{I}_1$ and $\bar{I}_2$. On the contrary, we observe that the term $\Psi^{(\bar{I}_1,\bar{I}_2)}$ remains consistently inactive.

\begin{figure}[h!]
    \centering
    \includegraphics[width=0.87\linewidth]{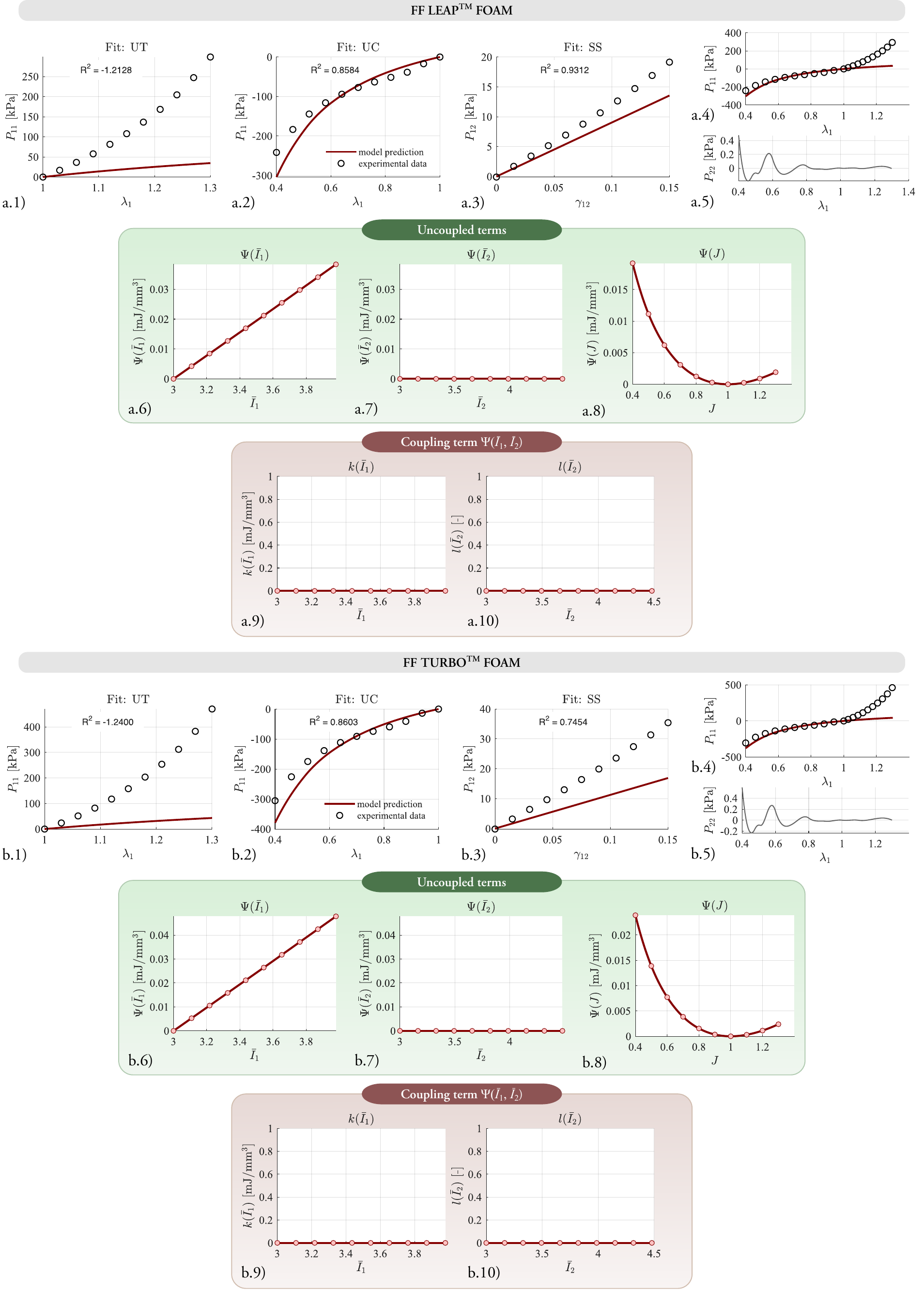}
    \caption{\textbf{Discovered hyperelastic strain-energy density functions with a mixed-invariant term $\Psi^{(\bar{I}_1,\bar{I}_2)}$ for FF LEAP$^\mathrm{TM}$ and FF TURBO$^\mathrm{TM}$ PLUS.}
(a,b.1--4) Model predictions for uniaxial tension (UT), uniaxial compression (UC), and simple shear (SS) deformation modes of the FF LEAP$^\mathrm{TM}$ and FF TURBO$^\mathrm{TM}$ foams. Nominal stress--stretch responses are shown alongside the experimental data used in the loss function.
(a,b.5) Predicted lateral stress response, for zero lateral stress condition enforced.
(a,b.6--10) Learned spline-based strain-energy density contributions.}
    \label{fig:results_I2I1_withoutothers}
\end{figure}

The findings in the previous sections already show that the uncoupled term $\Psi^{(\bar{I}_2)}$ is not really necessary to capture the extreme deformations and that the $\bar{I}_1$--$\bar{I}_2$--uncoupled terms $\Psi^{(\bar{I}_1,J)}$ and $\Psi^{(\bar{I}_2,J)}$ yield convincingly good predictions of the stress-strain curves across the three deformation modes. For the kinematics at hand (ultra-compressible foams under homogeneous deformation), additional results in \cref{fig:results_I2I1_withoutothers} confirm that a coupling between $\bar{I}_1$ and $\bar{I}_2$ is not useful: Both auxiliary spline functions $k^{(\bar{I}_1)}$ and $l^{(\bar{I}_2)}$ composing the term $\Psi^{(\bar{I}_1,\bar{I}_2)}$ are exactly zero (cf. subfigure a.9-10 for the FF LEAP$^\mathrm{TM}$ foam and subfigure b.9-10 for the FF TURBO$^\mathrm{TM}$ foam). Consequently, the results match those obtained in \cref{sec:only_uncoupled} for an ansatz with only uncoupled terms.

On a final note for this section, these results anticipate that a hypothetical threefold coupling term $\Psi^{(\bar{I}_1,\bar{I}_2,J)}$, consisting of three splines coupled via multiplicative decomposition, is not necessary. Therefore, we do not formulate such a term in the structure of our model (cf. \cref{eq:B-spline_energy}) but we acknowledge that it might become relevant for other kinematics in future investigations.

\subsection{Overparameterized, overly flexible strain-energy density functions lead to non-unique fits}

We finish with two overly rich assumptions: purely separable energy contributions with several coupling terms, i.e.,
{\setlength{\abovedisplayskip}{4pt}
\setlength{\abovedisplayshortskip}{4pt}
\begin{align}
     &\mathrm{Ansatz~1:} \quad && \Psi^{(\bar{I}_1,\bar{I}_2,J)} = \Psi^{(\bar{I}_1)} + \Psi^{(\bar{I}_2)} + \Psi^{(J)} + \Psi^{(\bar{I}_1,J)} + \Psi^{(\bar{I}_2,J)}; \nonumber \\
     &\mathrm{Ansatz~2:} \quad && \Psi^{(\bar{I}_1,\bar{I}_2,J)} = \Psi^{(\bar{I}_1)} + \Psi^{(\bar{I}_2)} + \Psi^{(J)} + \Psi^{(\bar{I}_1,J)} + \Psi^{(\bar{I}_2,J)} + \Psi^{(\bar{I}_1,\bar{I}_2)}. \nonumber
\end{align}}

The goal here is to intentionally promote non-uniqueness in spline-based energy representations. This contradicts conventional approaches where sparsity is encouraged through regularization of the energy and sparsity promoting techniques. We utilize spline-based strain-energy density functions as a data-adaptive tool to expose this issue and to illustrate the extent to which a hyperelastic energy function can be ambiguous for a multi-mode experimental dataset.\footnote{Note that we always refer to non-uniqueness across the terms added up to represent the strain-energy density function, and not to non-uniqueness in the calibration of the spline-based functions inside the individual coupling terms $\Psi^{(\bar{I}_1,J)}=g^{(J)}\,h^{(\bar{I}_1)}$, $\Psi^{(\bar{I}_2,J)}=i^{(J)}\,j^{(\bar{I}_2)}$, and $\Psi^{(\bar{I}_1,\bar{I}_2)}=k^{(\bar{I}_1)}\,l^{(\bar{I}_2)}$. Regarding the latter, we acknowledge that uniqueness of the auxiliary functions can be strictly enforced. However, we deem this unnecessary in the scope of the present work.} We argue that this deliberately non-standard perspective is important to highlight that multiple distinct representations can yield equally accurate predictions. Importantly, this behavior is not specific to our method but is a general challenge in material modeling. In this context, spline-based strain-energy density functions provide an effective tool for visualizing such non-uniqueness.

For the first representation with $\Psi^{(\bar{I}_1,J)}$ and $\Psi^{(\bar{I}_2,J)}$ coupling terms we present two solutions for each of the foams: \cref{fig:results_I1JI2J} and \cref{fig:results_I1JI2J_Bis1} in Appendix. 
The \textbf{first} solution (\cref{fig:results_I1JI2J}) leverages both $\bar{I}_1$--$J$ and $\bar{I}_2$--$J$ coupling terms to yield yet a good fitting 
(FF LEAP$^\mathrm{TM}$ foam: $\mathrm{R}^2=1.00$ for tension, $\mathrm{R}^2=1.00$ for compression, and $\mathrm{R}^2=1.00$ for shear; 
FF TURBO$^\mathrm{TM}$ foam: $\mathrm{R}^2=1.00$ for tension, $\mathrm{R}^2=0.98$ for compression, and $\mathrm{R}^2=0.97$ for shear). The $J$-dependent functions $g^{(J)}$ and $i^{(J)}$ are monotonous. 
The \textbf{second} solution (\cref{fig:results_I1JI2J_Bis1}) produces a similarly good fitting 
(FF LEAP$^\mathrm{TM}$ foam: $\mathrm{R}^2=1.00$ for tension, $\mathrm{R}^2=1.00$ for compression, and $\mathrm{R}^2=1.00$ for shear; 
FF TURBO$^\mathrm{TM}$ foam: $\mathrm{R}^2=1.00$ for tension, $\mathrm{R}^2=0.98$ for compression, and $\mathrm{R}^2=0.97$ for shear). For the TURBO$^\mathrm{TM}$ foam, the contribution of $\Psi^{(\bar{I}_2,J)}$ is almost zero (subfigure a.11-12).

For the second representation with $\Psi^{(\bar{I}_1,J)}$, $\Psi^{(\bar{I}_2,J)}$, and $\Psi^{(\bar{I}_1,\bar{I}_2)}$, we present two other solutions ---obtained with two different (random) initial guesses--- for each of the foams: \cref{fig:results_I1JI2JI1I2} and \cref{fig:results_I1JI2JI1I2_Bis1} in Appendix. Interestingly, we observe that the coupling term $\Psi^{(\bar{I}_1,\bar{I}_2)}$ (both its $k^{(\bar{I}_1)}$ and $l^{(\bar{I}_2)}$ spline-based components) is always identified as zero (subfigures a.9-10). Therefore, these results further illustrate the ambiguity associated with the terms $\Psi^{(\bar{I}_1,J)}$ and $\Psi^{(\bar{I}_2,J)}$, as discussed in the previous paragraphs. 
The first (or rather \textbf{third}) solution (\cref{fig:results_I1JI2JI1I2}) is interesting because it identifies a non-monotonous $g^{(J)}$ contribution (subfigure a.12) for the FF LEAP$^\mathrm{TM}$ foam and yet convincing results for both foams. 
The second (or rather \textbf{fourth}) solution (\cref{fig:results_I1JI2JI1I2_Bis1}) identifies a zero $h^{(\bar{I}_1)}$ function (subfigure a.11) for the FF LEAP$^\mathrm{TM}$ foam and combines a representation based on the uncoupled term $\Psi^{(\bar{I}_1)}$ and the coupling one $\Psi^{(\bar{I}_1,J)}$ (subfigures a.6, 13-14). For the FF TURBO$^\mathrm{TM}$ foam, the representation this time is dominated by $\bar{I}_2$. For the sake of brevity we refer directly to \cref{fig:results_I1JI2JI1I2} and \cref{fig:results_I1JI2JI1I2_Bis1} for the $\mathrm{R}^2$ values of these two last solutions.

\begin{figure}[H]
    \centering
    \includegraphics[width=0.95\linewidth]{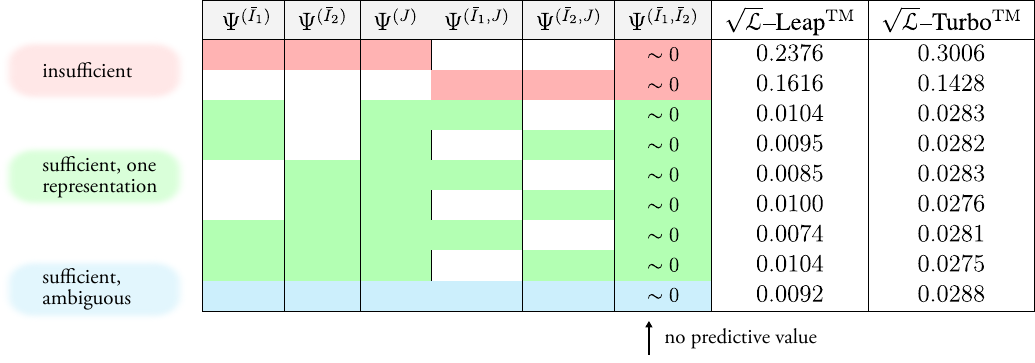}
    \caption{\textbf{Overview of the performance of the spline-based strain-energy density functions in capturing extreme compressible hyperelasticity.} 
The colored cells indicate that the contribution is active for a given ansatz. 
\textit{Red} denotes the inability of the ansatz to capture the combined uniaxial tension, compression, and simple shear behavior. For the first ansatz with the three uncoupled terms, the representation found by the optimizer is always the same, irrespective of the random initial guess. 
\textit{Green} denotes successful prediction with an unique representation found by the optimizer, independent of the random initial guess. 
\textit{Blue} indicates successful prediction; however, different representations are obtained when the calibration is performed with different initial guesses. 
The optimizer consistently suppresses the coupling term $\Psi(\bar{I}_1,\bar{I}_2)$, indicating that it provides no additional predictive value for the experimental data.
That a prediction is successful is assessed with the square root of the loss function $\sqrt{\mathcal{L}}$ being less than \qty{0.03}{}. 
Note that ambiguous solutions refer to non-uniqueness across different terms, and not to non-uniqueness arising from the calibration of the auxiliary functions within a single coupling term.
Furthermore, the results in this table reflect those obtained using the bespoke optimizer with random initial guesses; they do not necessarily constitute general evidence.}
    \label{fig:summarize}
\end{figure}

To summarize, the table in \cref{fig:summarize} provides an overview of how different assumptions for the strain-energy density function predict the experimental curves. It is clear that an ansatz consisting solely of uncoupled terms is not sufficient to capture the tension--compression asymmetry. Likewise, an ansatz consisting solely of coupling terms is also insufficient. To capture large volumetric deformations, it is necessary to combine coupling terms with uncoupled terms; more precisely, among the uncoupled terms, the purely volumetric contribution $\Psi^{(J)}$ is the only truly required one. The solution is unique when the ansatz includes a subset of terms, and is non-unique when all terms are activated. It is also interesting that the term $\Psi(\bar{I}_1,\bar{I}_2)$ does not contribute to predictive stress: The optimizer consistently iterates it to zero.

\begin{figure}[H]
    \centering
    \includegraphics[width=0.9\linewidth]{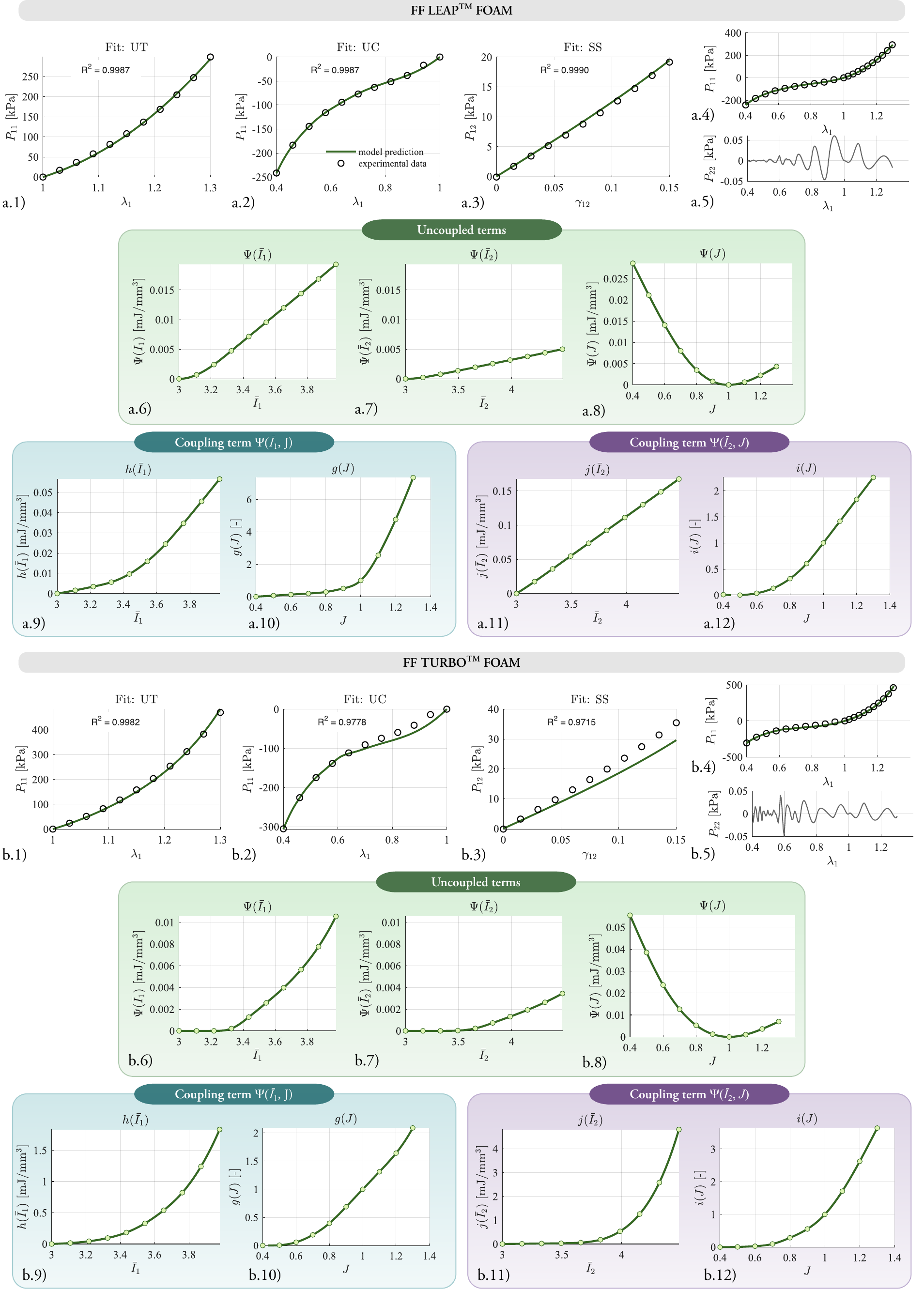}
    \caption{\textbf{Discovered hyperelastic strain-energy density functions with mixed-invariant terms $\Psi^{(\bar{I}_1,J)}$ and $\Psi^{(\bar{I}_2,J)}$ for FF LEAP$^\mathrm{TM}$ and FF TURBO$^\mathrm{TM}$ PLUS.}
(a,b.1--4) Model predictions for uniaxial tension (UT), uniaxial compression (UC), and simple shear (SS) deformation modes of the FF LEAP$^\mathrm{TM}$ and FF TURBO$^\mathrm{TM}$ foams. Nominal stress--stretch responses are shown alongside the experimental data used in the loss function.
(a,b.5) Predicted lateral stress response, for zero lateral stress condition enforced.
(a,b.6--12) Learned spline-based strain-energy density contributions.}
    \label{fig:results_I1JI2J}
\end{figure}

\begin{figure}[H]
    \centering
    \includegraphics[width=1\linewidth]{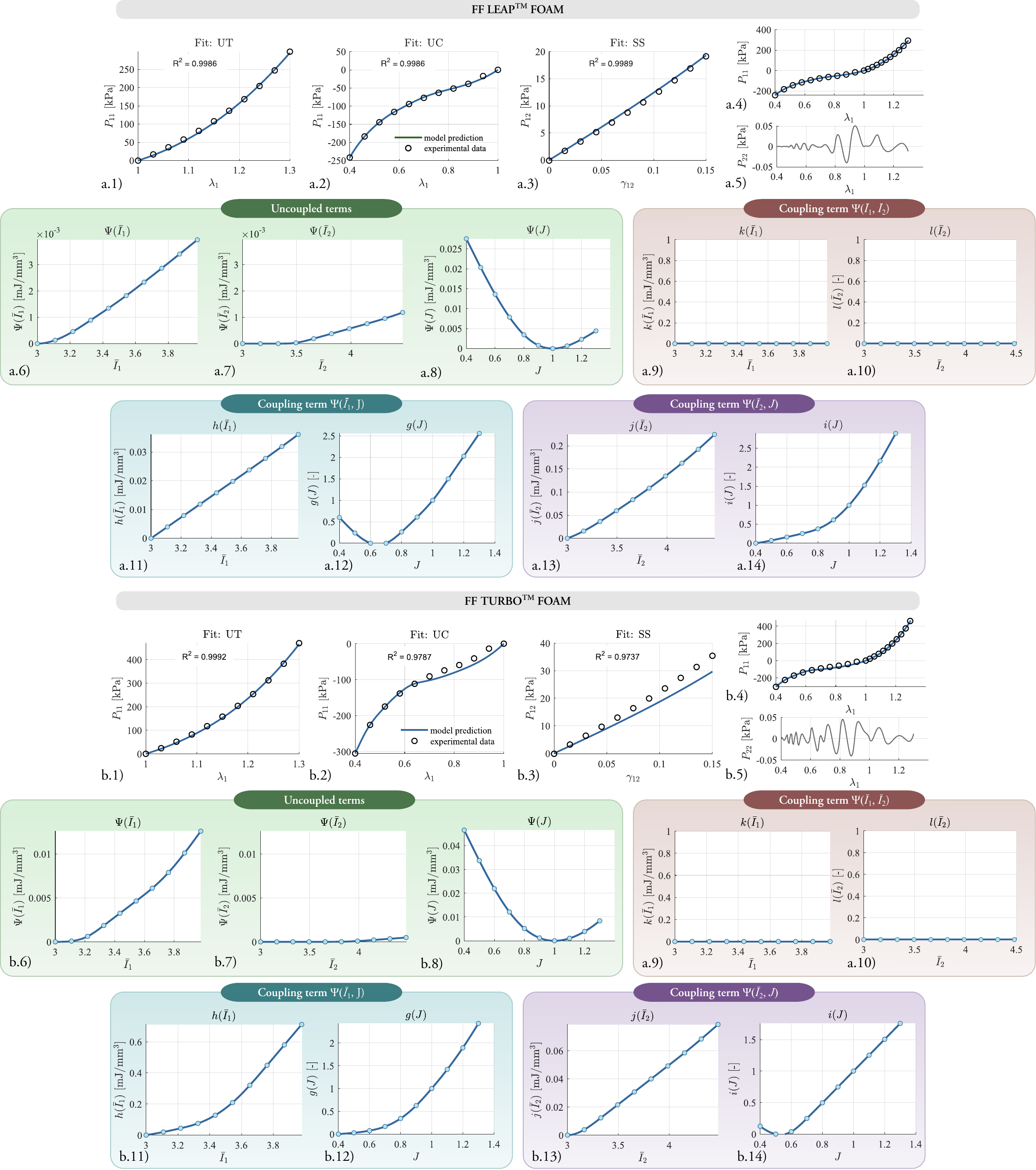}
    \caption{\textbf{Discovered hyperelastic strain-energy density functions with mixed-invariant terms $\Psi^{(\bar{I}_1,J)}$, $\Psi^{(\bar{I}_2,J)}$, and $\Psi^{(\bar{I}_1,\bar{I}_2)}$ for FF LEAP$^\mathrm{TM}$ and FF TURBO$^\mathrm{TM}$ PLUS.}
(a,b.1--4) Model predictions for uniaxial tension (UT), uniaxial compression (UC), and simple shear (SS) deformation modes of the FF LEAP$^\mathrm{TM}$ and FF TURBO$^\mathrm{TM}$ foams. Nominal stress--stretch responses are shown alongside the experimental data used in the loss function.
(a,b.5) Predicted lateral stress response, for zero lateral stress condition enforced.
(a,b.6--14) Learned spline-based strain-energy density contributions.
Note that $g(J)$ in (a.12) becomes negative between $J\in[0.6,0.75]$. This occurs because convexity and the simple bounds $\theta>0$ are enforced on the spline coefficients.}
    \label{fig:results_I1JI2JI1I2}
\end{figure}

\section{Discussion \& Outlook}
Our data-adaptive spline-based energy density functions have captured convincingly well the tension--compression asymmetry of foams with coupling terms $\Psi^{(\bar{I}_1,J)}$ and $\Psi^{(\bar{I}_2,J)}$. The use of splines provides high adaptivity in their shape and curvature. The coupling terms were expressed as combinations through multiplicative decomposition. This ansatz is particularly effective to form a bilinear structure that becomes linear upon alternate optimization; and it allows for ultra-fast training.

The reader might argue that ensuring uniqueness in parameter identification requires a richer dataset. To some extent, this is true: Characterizations in the literature reach stretches $\lambda>5$ \cite{Treloar1944Stress-strainDeformation}, which is above the maximum stretches of the dataset that we use ($\lambda_1=0.4$ for compression, $\lambda_1=1.3$ for tension, and $\gamma_{12}=0.15$ for shear). This idea has motivated geometry-driven experimental protocols that deliberately introduce controlled heterogeneity in order to probe localization and multiaxial stress states. Such approaches are often embedded in unsupervised learning frameworks, where a finite element boundary value problem links the experimental observations and the constitutive model. 
Nevertheless, recent studies have shown that even rich deformation fields can fail to uniquely identify a “best” model among many competing candidates. In other words, multiple constitutive models can provide similarly accurate fits while being fundamentally different in structure, at least for yet not-rich enough heterogeneous displacement or strain fields. Importantly, these challenges are not specific to spline-based constitutive modeling but extend to basically all traditional and modern approaches of constitutive modeling.

A particularly suitable framework to analyze the non-uniqueness mater is Material Fingerprinting. It enables a systematic quantification of the “goodness” of the fit in a wide range of models through similarity measures. The work of the authors in \cite{Flaschel2026UnsupervisedMeasurements} refers to this limitation: Several distinct hyperelastic models capture the same experimental data with comparable precision, in spite of using heterogeneous measurements. 

The question so far has been which experimental data are necessary to calibrate unique strain-energy density functions. However, we recognize that certain extreme deformation modes may not arise in real applications of soft solids, or may simply be unattainable because the material fails before reaching them. Therefore, we propose a perspective that accepts slightly incomplete, yet realistic datasets, and thus explicitly addresses model non-uniqueness, as both pertinent and unavoidable rather than something to be dismissed or to hide. Strategies to diagnose and \textit{control} (and not \textit{hide}) this degeneracy include regularizations for sparsity of terms. However, when two terms capture the behavior equally well for a given (often limited) kinematic exploration (cf., our coupling terms $\Psi^{(\bar{I}_1,J)}$ and $\Psi^{(\bar{I}_2,J)}$ for ultra-compressible foams under homogeneous deformation), the term that should prevail may just be a matter of taste.

\section*{Code Availability}
\noindent The \textsc{Matlab} code generated for the calibration of the spline-based strain-energy density functions is publicly available in
\url{https://github.com/MiguelAngelMorenoMateos/compressible-spline-hyperelasticity}.

\section*{Acknowledgments}
\noindent The authors acknowledge support from the European Research Council (ERC) under the Horizon Europe program upon the two following grants: Grant-No. 101052785, project: SoftFrac; and Grant-No. 101141626, project: DISCOVER. Funded by the European Union. Views and opinions expressed are however those of the authors only and do not necessarily reflect those of the European Union or the European Research Council Executive Agency. Neither the European Union nor the granting authority can be held responsible for them. Furthermore, Miguel Angel Moreno-Mateos gratefully acknowledges support from the FAU Competence Center Engineering of Advanced Materials (FAU EAM) through an EAM Starting Grant. Miguel Angel Moreno-Mateos also acknowledges the members of the Living Matter Lab for the insightful discussions during his research stay at Stanford University in February 2026.
\begin{figure}[H]
\includegraphics[width=0.3\textwidth]{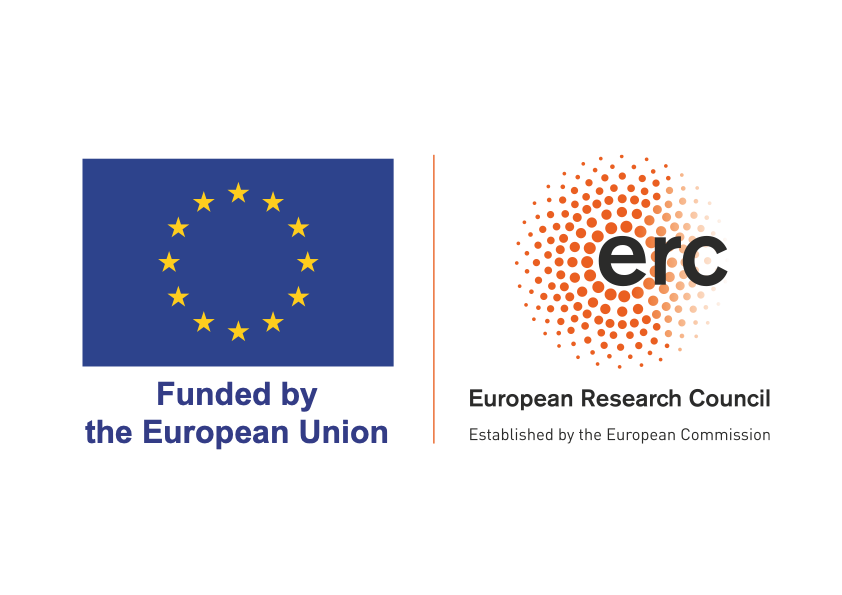}
\end{figure}

\section*{Competing Interests}
\noindent The Authors declare no Competing Financial or Non-Financial Interests.

\appendix

\section{Linear inequality constraints}\label{sec:linear_constraints}

The spline-based representation of the strain-energy density function enables the enforcement of convexity and monotonicity constraints through linear inequalities on the spline coefficients. These constraints ensure physically meaningful behavior such as material stability and non-negative stiffness.

To this end, we first need to relate the interpolation points to the control points. The energy contributions can be expressed with B-spline basis functions and respective control points (in the spirit of \cref{eq:univariate_splines,eq:coupling_terms}) or as a linear combination of the interpolation points ($\boldsymbol{\theta}$) and basis functions tied to such interpolation points (cf. \cref{eq:B-spline_interp_basis}).

From \cref{eq:B-spline_interp_basis}, the derivatives of the parameter sensitivity splines read

\begin{equation}
    \Psi'(x) = \sum_{p=1}^{P} \theta_p N_p'(x) \quad \mathrm{and} \quad \Psi''(x) = \sum_{p=1}^{P} \theta_p N_p''(x),
\end{equation}

\noindent for $P$ the number of optimization parameters.

As a consequence, any condition on the derivatives remains linear on the optimization parameters (interpolation values) $\boldsymbol{\theta}$. 

A linear mapping between the B-spline basis functions $\phi_i$ and the parameter sensitivity splines can be then defined as

\begin{equation}
    N_p(x)=\sum_{i=1}^P \hat{c}_{pi} \phi_i(x).
\end{equation}

The matrix $\mathbf{C}_\mathrm{ineq}$ is defined as the collection of all $c_{pi}$ coefficients, i.e.,
\begin{equation}
    \mathbf{C}_\mathrm{ineq} = 
    \begin{bmatrix}
        c_{11} & \dots & c_{1P}\\
        \vdots &       & \vdots\\
        c_{P1} & \dots & c_{PP}
    \end{bmatrix},
\end{equation}

\noindent where each row is related to one sensitivity spline $N_p(x)$ and each column, to one B-spline basis function. The matrix $\mathbf{C}$ relates the B-spline basis and the parameter sensitivity splines. Therefore, we can write

\begin{equation}
    \Psi(x) = \sum_{i=1}^P \left[ \sum_{p=1}^P \theta_p \hat{c}_{pi} \right] \phi_i(x), 
\end{equation}

\noindent where $\sum_{i=1}^P \hat{c}_{pi}$ becomes the coefficient (control point) of the B-spline basis function $\phi$. This forms the $i$-th row of the ineq. constraint matrix.

The derivative reads

\begin{equation}
    \Psi'(x)  =
    \sum_{i=1}^{P-1} \left[ \sum_{p=1}^P \theta_p \hat{c}_{pi}^{(1)} \right] \phi_i(x).
\end{equation}

Since the B-spline basis $\phi_i$ are non-negative and form a partition of unity, a sufficient condition for 
\begin{equation}
    \Psi'(x)\geq 0  \quad \forall \, x ,
\end{equation}

\noindent is that the coefficients are non-negative,
\begin{equation}
    \sum_{p=1}^P \theta_p \hat{c}_{pi} \geq 0,
\end{equation}

\noindent which can be rewriten in matrix form as 
\begin{equation}
    -\mathbf{C}_\mathrm{ineq}^{(1)}\boldsymbol{\theta} \leq 0.
\end{equation}

The similar reasoning applies to the second derivative to enforce convexity,
\begin{equation}
    \Psi''(x)\geq 0  \quad \forall \, x ,
\end{equation}

\noindent enforced requiring non-negativity of the related coefficients, i.e.,
\begin{equation}
    -\mathbf{C}_\mathrm{ineq}^{(2)}\boldsymbol{\theta} \leq 0.
\end{equation}

These constraints can be generalized to all energy contributions in the constitutive model. In particular, monotonicity and convexity are enforced on selected terms, while convexity-only conditions are imposed on volumetric and coupling contributions. This leads to the following global set of inequality constraints,
\begin{align}\label{eq:optim_constraints}
& \frac{\partial \phi}{\partial x} \ge 0, \quad
  \frac{\partial^2 \phi}{\partial x^2} \ge 0,
  \quad
  &&\forall \, \phi \in \{\Psi^{(\bar{I}_1)},\,\Psi^{(\bar{I}_2)},\,h^{(\bar{I}_1)},\,j^{(\bar{I}_2)},\,k^{(\bar{I}_1)},\,l^{(\bar{I}_2)}\}, \nonumber \\
& \frac{\partial^2 \phi}{\partial x^2} \ge 0,
  \quad
  &&\forall \, \phi \in \{\Psi^{(J)},\,g^{(J)},\,i^{(J)}\}.
\end{align}

Collecting all linear inequalities, the constraint system can finally be arranged in the matrices
\begin{equation}
\mathbf{A}_{\mathrm{ineq}}=
    \begin{bmatrix}
-\mathbf{C}_\mathrm{ineq}^{(1)}\\
-\mathbf{C}_\mathrm{ineq}^{(2)}
\end{bmatrix}.
\end{equation}

The global inequality constraint takes the form
\begin{equation}
\mathbf{A}_{\mathrm{ineq}} \boldsymbol{\theta} \le \mathbf{b}_{\mathrm{ineq}}.
\end{equation}

\section{Additional results for non-unique energies and stochastic initializations}

\begin{figure}[H]
    \centering
    \includegraphics[width=1\linewidth]{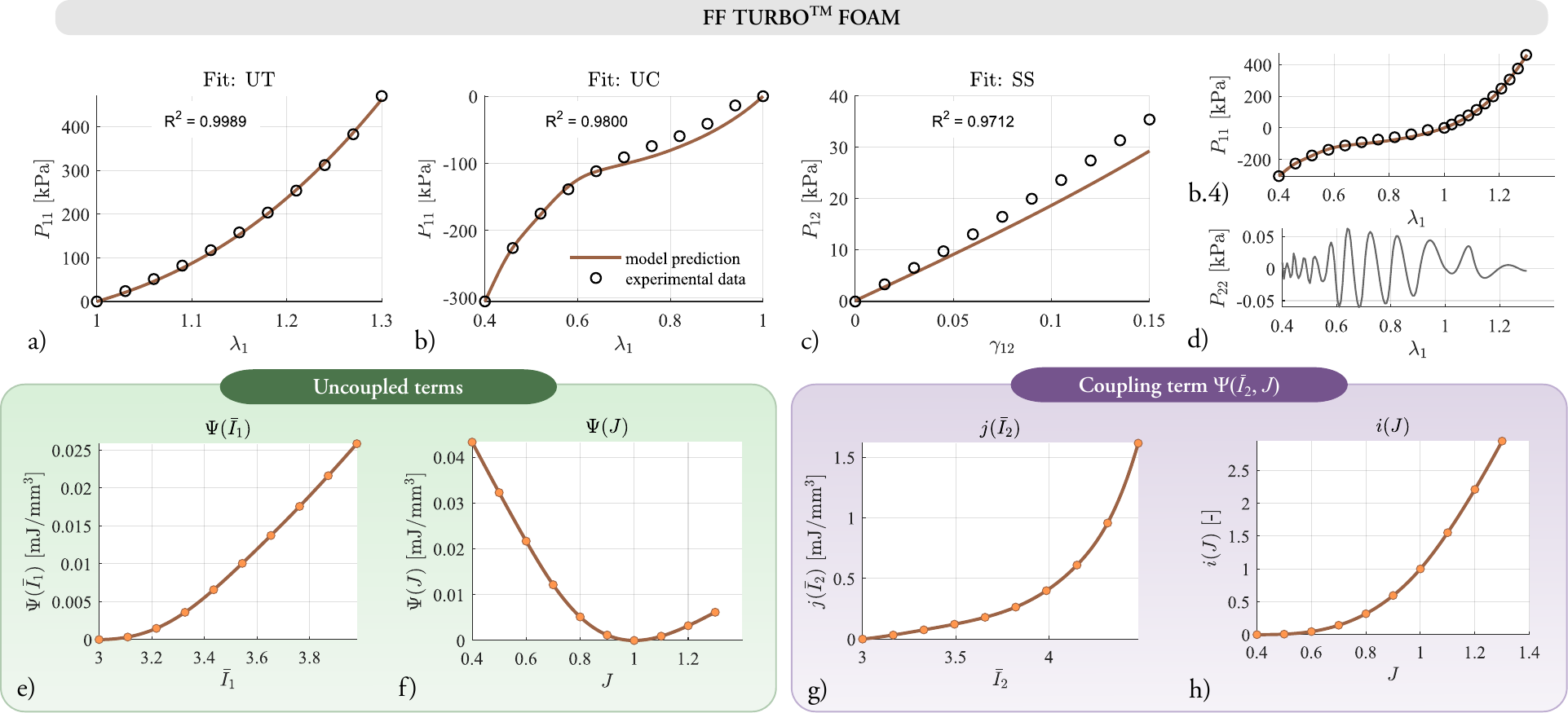}
    \caption{\textbf{Additional discovered hyperelastic strain-energy density functions with a mixed-invariant term $\Psi(\bar{I}_2,J)$ and without the uncoupled term $\Psi^{(\bar{I}_2)}$ for FF TURBO$^\mathrm{TM}$ PLUS.}
(a,b.1--4) Model predictions for uniaxial tension (UT), uniaxial compression (UC), and simple shear (SS) deformation modes of the FF LEAP$^\mathrm{TM}$ and FF TURBO$^\mathrm{TM}$ foams. Nominal stress--stretch responses are shown alongside the experimental data used in the loss function.
(a,b.5) Predicted lateral stress response, for zero lateral stress condition enforced.
(a,b.6--9) Learned spline-based strain-energy density contributions.}
    \label{fig:results_I2J_withoutI2}
\end{figure}

\begin{figure}[H]
    \centering
    \includegraphics[width=0.9\linewidth]{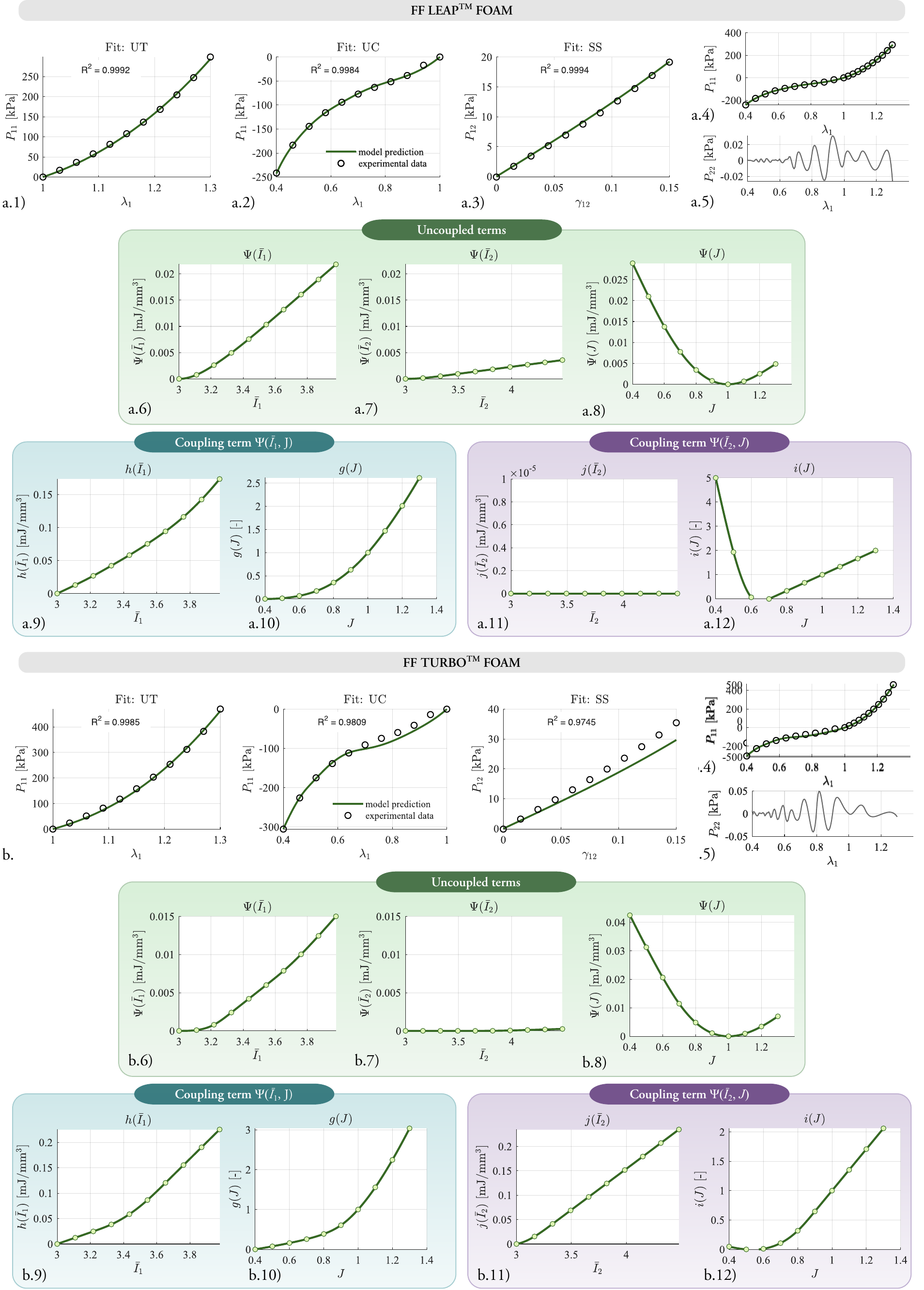}
    \caption{\textbf{Additional discovered hyperelastic strain-energy density functions for a different stochastic initial guess, with mixed-invariant terms $\Psi^{(\bar{I}_1,J)}$ and $\Psi^{(\bar{I}_2,J)}$ for FF LEAP$^\mathrm{TM}$ and FF TURBO$^\mathrm{TM}$ PLUS.}
(a,b.1--4) Model predictions for uniaxial tension (UT), uniaxial compression (UC), and simple shear (SS) deformation modes of the FF LEAP$^\mathrm{TM}$ and FF TURBO$^\mathrm{TM}$ foams. Nominal stress--stretch responses are shown alongside the experimental data used in the loss function.
(a,b.5) Predicted lateral stress response, for zero lateral stress condition enforced.
(a,b.6--12) Learned spline-based strain-energy density contributions.}    
    \label{fig:results_I1JI2J_Bis1}
\end{figure}

\begin{figure}[H]
    \centering
    \includegraphics[width=1\linewidth]{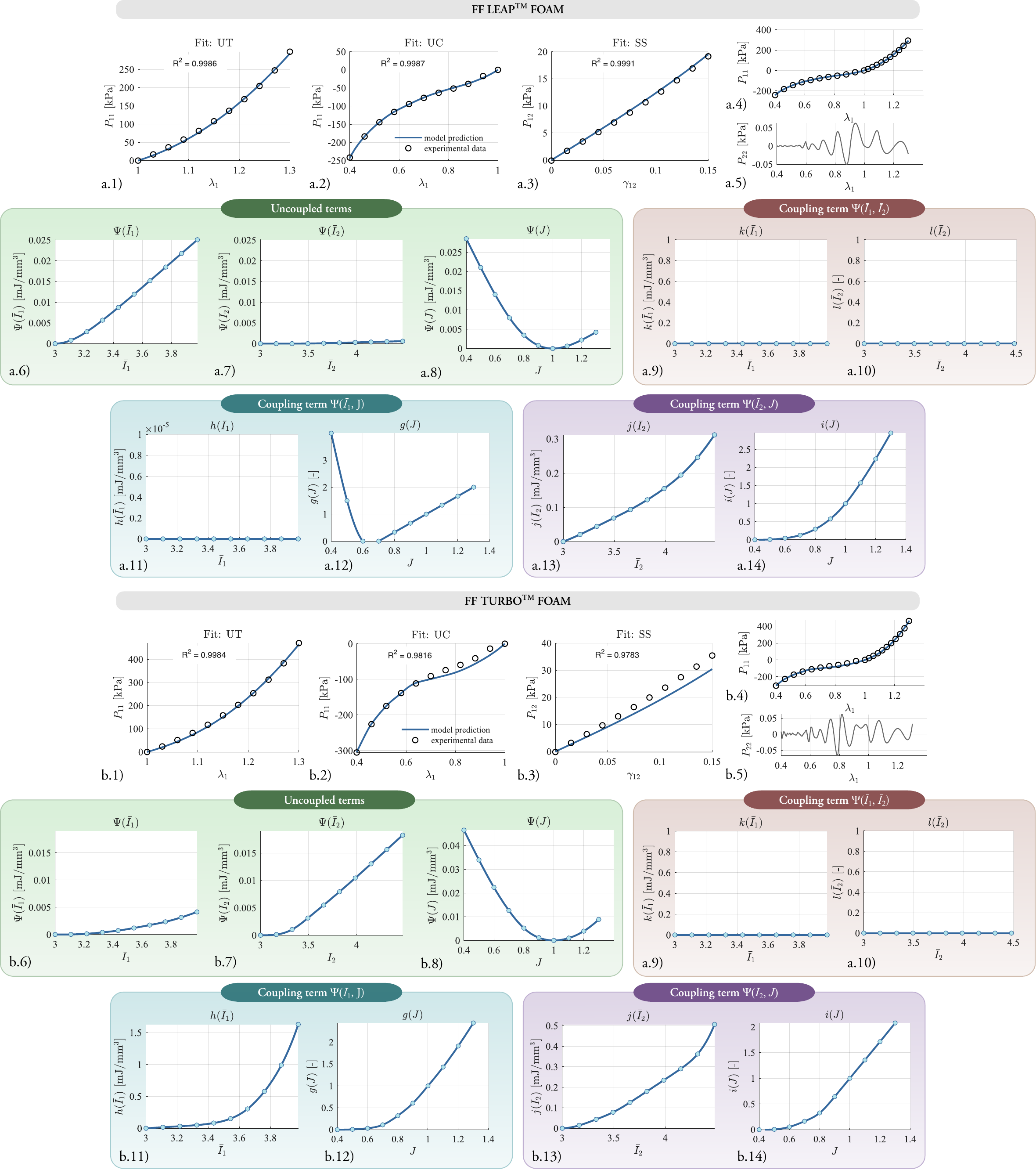}
    \caption{\textbf{Additional discovered hyperelastic strain-energy density functions for a different stochastic initial guess, with mixed-invariant terms $\Psi^{(\bar{I}_1,J)}$, $\Psi^{(\bar{I}_2,J)}$, and $\Psi^{(\bar{I}_1,\bar{I}_2)}$ for FF LEAP$^\mathrm{TM}$ and FF TURBO$^\mathrm{TM}$ PLUS.}
(a,b.1--4) Model predictions for uniaxial tension (UT), uniaxial compression (UC), and simple shear (SS) deformation modes of the FF LEAP$^\mathrm{TM}$ and FF TURBO$^\mathrm{TM}$ foams. Nominal stress--stretch responses are shown alongside the experimental data used in the loss function.
(a,b.5) Predicted lateral stress response, for zero lateral stress condition enforced.
(a,b.6--12) Learned spline-based strain-energy density contributions.}
    \label{fig:results_I1JI2JI1I2_Bis1}
\end{figure}

We present a second set of complementary results to illustrate the non-uniqueness in representing hyperelastic compressible constitutive behaviors with data-adaptive spline-based energy density functions. This includes results for the following models:
{\setlength{\abovedisplayskip}{4pt}
\setlength{\abovedisplayshortskip}{4pt}
\begin{align}
    &\mathrm{Ansatz~1,~\cref{fig:results_I2J_withoutI2}:} \quad && \Psi^{(\bar{I}_1,\bar{I}_2,J)} = \Psi^{(\bar{I}_1)} +  \Psi^{(J)} + \Psi^{(\bar{I}_2,J)}; \nonumber \\
     &\mathrm{Ansatz~2,~\cref{fig:results_I1JI2J_Bis1}:} \quad && \Psi^{(\bar{I}_1,\bar{I}_2,J)} = \Psi^{(\bar{I}_1)} + \Psi^{(\bar{I}_2)} + \Psi^{(J)} + \Psi^{(\bar{I}_1,J)} + \Psi^{(\bar{I}_2,J)}; \nonumber \\
     &\mathrm{Ansatz~3,~\cref{fig:results_I1JI2JI1I2_Bis1}:} \quad && \Psi^{(\bar{I}_1,\bar{I}_2,J)} = \Psi^{(\bar{I}_1)} + \Psi^{(\bar{I}_2)} + \Psi^{(J)} + \Psi^{(\bar{I}_1,J)} + \Psi^{(\bar{I}_2,J)} + \Psi^{(\bar{I}_1,\bar{I}_2)}. \nonumber
\end{align}}

\newpage

\bibliographystyle{naturemag}
\newpage

\addcontentsline{toc}{section}{References}
%\bibliography{bibliography}

\end{document}